\documentclass[a4paper,fleqn]{cas-sc}

\usepackage[authoryear]{natbib}
\usepackage[ansinew]{inputenc}
\usepackage[bbgreekl]{mathbbol}
\usepackage{amsmath,amssymb,amsthm}
\usepackage{amsfonts}
\usepackage{bm}
\usepackage{bbm}
\usepackage{graphicx} % Required for inserting images
\usepackage{subcaption} % Required for subfigure environment
\usepackage{textgreek}
\AtBeginDocument{
  \let\uprho\varrho
}
\newcommand{\bolduppsi}{ \text{\ooalign{\hfil\textpsi\hfil\cr\hfil\raise0.1ex\hbox{\textpsi}\hfil\cr}}
}
\newcommand{\boldupphi}{ \text{\ooalign{\hfil\textpsi\hfil\cr\hfil\raise0.1ex\hbox{\textphi}\hfil\cr}}
}
\newcommand{\bolduprho}{ \text{\ooalign{\hfil\textpsi\hfil\cr\hfil\raise0.1ex\hbox{\textrho}\hfil\cr}}
}
\newcommand{\fracp}[2]{\frac{\partial #1}{\partial #2}}
\newcommand{\fract}[2]{\frac{\mathrm{d} #1}{\mathrm{d} #2}}
\newcommand{\dd}{\mathop{}\!\mathrm{d}} %for integrals

\newcommand{\ii}{\mathop{}\!\mathrm{i}}
\newcommand{\sinc}{\mathop{}\!\mathrm{sinc}}

\newcommand{\half}{\frac{1}{2}}
\newcommand\bb{\boldsymbol{b}}
\newcommand\bk{\boldsymbol{k}}
\newcommand\bX{\boldsymbol{X}}

\newcommand\bV{\boldsymbol{V}}

\newcommand\bx{\boldsymbol{x}}
\newcommand\bv{\boldsymbol{v}}
\newcommand\bE{\boldsymbol{E}}
\newcommand\bB{\boldsymbol{B}}
\newcommand\bA{\boldsymbol{A}}
\newcommand\bJ{\boldsymbol{J}}

\newcommand\bD{\boldsymbol{D}}
\newcommand\bH{\boldsymbol{H}}
\newcommand\bK{\boldsymbol{K}}
\newcommand\bL{\boldsymbol{L}}
\newcommand\bM{\boldsymbol{M}}
\newcommand\bN{\boldsymbol{N}}

\newcommand\cR{\mathcal{R}}
\newcommand\cC{\mathcal{C}}

\newcommand{\arr}[1]{{\bm{\mathsf{#1}}}}

\newcommand\arrA{{\bm{\mathsf{A}}}}
\newcommand\arrB{{\bm{\mathsf{B}}}}

\newcommand\arrE{{\bm{\mathsf{E}}}}

\newcommand\arrJ{{\bm{\mathsf{J}}}}

\def\matd{\mathbb{d}}

\def\matC{\mathbb{C}}
\def\matD{\mathbb{D}}

\def\matG{\mathbb{G}}
\def\matH{\mathbb{H}}

\def\matI{\mathbb{I}}

\def\matO{\mathbb{O}}

\begin{document}
\let\WriteBookmarks\relax
\def\floatpagepagefraction{1}
\def\textpagefraction{.001}
\shorttitle{Geometric discretization of quasineutral hybrid drift-kinetic model}
\shortauthors{N. Narechania et~al.}
%\begin{frontmatter}

\title [mode = title]{Geometric numerical discretization of a quasineutral hybrid model of drift-kinetic electrons and fully kinetic ions}                      
%\tnotemark[1,2]

%\tnotetext[1]{This document is the results of the research
%   project funded by the National Science Foundation.}

%\tnotetext[2]{The second title footnote which is a longer text matter
%   to fill through the whole text width and overflow into
%   another line in the footnotes area of the first page.}

\author[1]{Nishant Narechania}[orcid=0009-0007-1249-0843]
\cortext[cor1]{Corresponding author}
\ead{nishant.narechania@ipp.mpg.de}
\cormark[1]

\author[1]{Guo Meng}[orcid=0000-0002-0969-733X]
\author[1]{Emil Poulsen}[orcid=0000-0003-3826-5226]
\author[1,2]{Eric Sonnendr\"ucker}[orcid=0000-0002-8340-7230]

\affiliation[1]{organization={Numerical Methods in Plasma Physics, Max Planck Institute for Plasma Physics},
                addressline={Boltzmannstra{\ss}e 2}, 
                city={Garching},
               citysep={},
                postcode={85748}, 
                state={Bavaria},
                country={Germany}}

\affiliation[2]{organization={School of Computation, Information and Technology, Technical University of Munich},
                addressline={Boltzmannstra{\ss}e 3}, 
                city={Garching},
               citysep={},
                postcode={85748}, 
                state={Bavaria},
                country={Germany}}

\begin{abstract}
 We extend the geometric electromagnetic particle-in-cell (PIC) framework, \texttt{GEMPICX}, to solve the quasineutral hybrid Vlasov-Maxwell equations with drift-kinetic electrons and fully kinetic ions. A structure-preserving finite difference method that employs dual grids is used. The discrete action principle for the hybrid model is derived, using the dual nature of the grids. The dynamical system for this hybrid quasineutral model does not explicitly involve the temporal evolution term for the electric field. A curl-curl equation is therefore used to implicitly obtain the component of the electric field that is parallel to the background magnetic field, at every timestep. The perpendicular component of the electric field is obtained using the quasineutral Amp\`ere's equation without the displacement current, combined with the definition of the current in the drift-kinetic model. The discretized versions of the electric field equations are large, sparse linear systems. A fully explicit time-stepping scheme as well as two implicit-explicit (IMEX) schemes are tested. The numerical model is validated by verifying the various waves obtained from the dispersion relation.
\end{abstract}

\begin{keywords}
structure-preserving methods \sep quasineutral models \sep hybrid PIC methods \sep drift-kinetics
\end{keywords}

\maketitle

%%%%%%%%%%%%%%%%%%%%%%%%%%%%%%%%%%%%%%%%%%%%%%%%%%%%%%%%%
\section{Introduction}\label{sec:intro}

Hybrid kinetic models, that are a combination of fully kinetic and reduced models, are fundamental to modern computational plasma physics \citep{lipatov2002hybrid}. They present an effective balance between computational efficiency and physical accuracy in applications where kinetic effects are essential but a full kinetic treatment of all species is prohibitively expensive. In many plasma physics applications, the large mass disparity between ions and electrons leads to their dynamics occurring over significantly different spatial and temporal scales \citep{hazeltine2003plasma}. A widely adopted hybrid modelling approach therefore is to treat ions kinetically, while approximating electrons using reduced models. Quasineutral hybrid models are especially appealing in the context of computational efficiency. In applications with physical dimensions greatly exceeding the Debye length, the quasineutrality approximation is generally valid. By neglecting the displacement current in the Amp\`ere equation, quasineutral models filter out the small length scales associated with charge separation phenomena, and also eliminate high-frequency electromagnetic wave propagation from the full Vlasov-Maxwell system. They therefore significantly relax the stringent time-step restrictions tied to the full Vlasov-Maxwell system, while still maintaining key kinetic effects. In the reduced drift-kinetic electron model, the electron gyromotion is modelled as a limiting case of gyrokinetic theory in the small Larmor radius limit \citep{burby2019gauge-free}. This removes the description of phenomena occurring at electron gyromotion time and length scales. Therefore, in hybrid drift-kinetic quasineutral models, with kinetic ions and drift-kinetic electrons, the time-step restriction is relaxed even further by the use of reduced models for electron motion coupled with quasineutrality.

Hybrid plasma models span a broad class of approaches that combine kinetic and reduced descriptions for different species. On one hand, hybrid Vlasov models treat ions kinetically via a phase-space distribution while modelling electrons as a fluid, providing a noise-free but computationally expensive alternative to particle methods \citep{valentini2007753}. On the other hand, the more widely used hybrid particle-in-cell (PIC) models represent ions with particles and electrons as a fluid, as established in earlier works \citep{winske1985hybrid,matthews1994current,lipatov2002hybrid}. These ion-kinetic/electron-fluid models remain the standard in many applications and have seen continued development in recent years, including large-scale and application-oriented codes such as \texttt{AHKASH} \citep{chirakkara}, as well as improved hybrid PIC formulations \citep{jiao2025, wu2024, joshi2023fluid}, often employing simplified closures such as adiabatic electrons, where the electron response is assumed to follow a prescribed equation of state rather than being evolved dynamically.

Beyond fluid closures, there is a growing class of hybrid models with kinetic ions and reduced kinetic descriptions for electrons, such as gyrokinetic or drift-kinetic descriptions, providing a balance between fluid and kinetic descriptions of electrons. Such models, which are also the focus of this work, are rooted in the theoretical framework of \citet{brizard2007foundations}, and are particularly well suited to strongly magnetized plasmas. In addition, hybrid formulations based directly on the electric and magnetic fields, avoiding the use of electromagnetic potentials, have been developed in recent years \citep{chen2009particle, chen2019new}. Such models have been explored in a variety of contexts. For example, \citet{tyushev2025} performed drift-kinetic PIC simulations of magnetic mirror configurations and transport problems. \citet{glinskiy2024} developed semi-implicit drift-kinetic PIC schemes for plasma confinement and heating processes. \citet{seo2021} developed a hybrid drift-kinetic PIC model for studying plasma turbulence in a cylindrical tokamak geometry. These developments provide the conceptual and numerical foundation for the present work, which combines kinetic ions with drift-kinetic electrons in a structure-preserving quasineutral PIC framework.

Despite their advantages, the numerical solution of hybrid quasineutral models presents substantial challenges and remains an active field of research \citep{hockney2021computer, birdsall2018plasma}. Conventional PIC methods usually suffer from long-term numerical instabilities, violation of conservation laws, and unphysical energy growth. These issues are caused due to the lack of compatibility between discrete field solvers and particle updates, and to the breakdown of geometric properties of the continuous system at the discrete level. Structure-preserving methods that preserve certain geometric structures of the physical system of governing equations, such as conservation laws, gauge symmetry, and the Gauss laws, have emerged as a reliable approach to address these difficulties. For example, various researchers have developed discrete energy-conserving semi-implicit \citep{chen2011energy, chen2015multi} and fully implicit \citep{markidis2011energy, lapenta2017exactly} PIC schemes. A subset of structure-preserving methods are methods that exactly preserve discretized equivalents of invariants by discretizing the action principle or Hamiltonian structure of the governing equations. Recently, significant progress has been made in the development of structure-preserving PIC schemes, including variational integrators \citep{squire2012geometric, marsden2001discrete}, finite element exterior calculus (FEEC) approaches \citep{arnold2018finite}, and compatible discretizations based on the de Rham complex \citep{bossavit1998computational, hiptmair2002finite, tronci2014hybrid}. 

Methods based on the de Rham complex ensure an exact preservation of the Gauss laws after discretization. One such method, namely the finite element geometric electromagnetic PIC (\texttt{GEMPIC}) method, was developed  based on the Vlasov-Maxwell system's intrinsic Hamiltonian structure \citep{kraus2017gempic, campos2022variational, kormann2024}. This method conserved discretized versions of the Gauss laws, total energy, the Poisson structure and Casimir invariants. This method was then also extended to a discretization based on structure-preserving mimetic finite differences, now known as \texttt{GEMPICX} \citep{kormann2024}. Mimetic discretization methods use operators designed to preserve key identities from vector calculus within a discrete framework \citep{bochev2006principles}. In this setting, unknown quantities are expressed as values associated with points, as well as integrals over edges, faces, and volumes, forming a discrete analogue of the de Rham complex. More recently, \citet{meng2025} extended this framework to the drift-kinetic (DK) model and to hybrid models that combine drift-kinetic electrons with fully kinetic ions. Similarly, \citet{narechania2026FKQN} extended it to the quasineutral Vlasov-Maxwell model with fully kinetic descriptions of both species.

In this work, we extend \texttt{GEMPICX} to solve the quasineutral hybrid drift-kinetic model with drift-kinetic electrons and kinetic ions, using a structure-preserving discretization. 
This model eliminates fast light waves, thereby enabling efficient and realistic studies of Alfv\'en wave and ion cyclotron dynamics; the model formulation and its physical applications are described in a companion paper \citep{meng2026QN-DeFi}.
We start with the Lagrangian from \citet{burby2019gauge-free} and propose a discretized Lagrangian based on mimetic finite differences using dual grids. Discretized field equations and equations of motion for particles are obtained from the discretized Lagrangian using the variational principle, ensuring consistency with the de Rham geometric structure of the continuous equations. Particle-field coupling is achieved using spline-based interpolation consistent with this discretization. A key aspect of the proposed formulation is the decomposition of the electric field into components parallel and perpendicular to the background magnetic field. The removal of the displacement current eliminates the standard evolution equation for the electric field, requiring alternative formulations to determine it consistently. The perpendicular component is obtained from a modified quasineutral Amp\`ere equation, while the parallel component is determined from a curl-curl equation derived from Faraday's law and current evolution. These formulations lead to large, sparse linear systems, whose efficient treatment is essential for practical simulations. We therefore investigate both fully explicit and implicit-explicit (IMEX) time integration schemes, and analyze their stability properties using a cold plasma model. We validate the numerical model by examining the wave spectra of quasineutral plasmas, in a quasi-1D setting under periodic boundary conditions. Additionally, an ion cyclotron wave simulation with periodic boundaries is performed to demonstrate the capability of the numerical method to model a higher order process such as damping.

The organization of this paper is as follows: Section \ref{sec:hybriddkqnmodel} describes the quasineutral hybrid model with drift-kinetic electrons and fully kinetic ions, including the action principle, governing equations, and the electric field equations. Section \ref{sec:fielddiscretization} describes the structure-preserving dual-mesh approach used for discretization of fields. Section \ref{sec:particlediscretization} describes the coupling between particles and discretized field variables. Section \ref{sec:discreteqnvmmodel} describes the semi-discrete equivalents of the action principle, governing equations, and electric field equations. Section \ref{sec:stabilityanalysis} derives the time-step criterion for stability of the numerical time-stepping algorithm, using a cold plasma model. Section \ref{sec:timestepping} details the time-stepping schemes used for solving the semi-discrete system of equations. The dispersion relation and the various eigenmodes generated are discussed in Section \ref{sec:dispersionrelation}. Section \ref{sec:tests} describes numerical simulations with periodic boundary conditions performed to validate this numerical model. The parallelization strategy is described in Section \ref{sec:parallelization}. We finally summarize this work, draw conclusions and discuss future directions in Section \ref{sec:conclusions}.
%%%%%%%%%%%%%%%%%%%%%%%%%%%%%%%%%%%%%%%%%%%%%%%%%%%%%%%%%

%%%%%%%%%%%%%%%%%%%%%%%%%%%%%%%%%%%%%%%%%%%%%%%%%%%%%%%%%
\section{The quasineutral hybrid model with drift-kinetic electrons and fully kinetic ions}\label{sec:hybriddkqnmodel}
In this section, we describe the Lagrangian for the hybrid drift-kinetic quasineutral model, comprising terms associated with the electron kinetic energy, the ion kinetic energy and the magnetic field energy. We then obtain the governing equations from this action. Finally, the electric field equations are obtained from the governing equations. 

\subsection{Action principle}\label{sec:action}
We first describe the Lagrangian and action principle for the electrons in our hybrid model, that are described using the drift-kinetic model. The drift-kinetic model is obtained from the gyrokinetic model described by \citet{burby2019gauge-free}, by taking the zero Larmor radius limit. Let $\bB_{eq}$ be the time-invariant, equilibrium background magnetic field acting on the plasma and $\bb_{eq} = \bB_{eq}/|\bB_{eq}|$ its associated unit vector. $\bE$ and $\bB$ denote the perturbed electric and magnetic fields, respectively. Their components perpendicular to $\bB_{eq}$ are therefore given by the vectors $\bB\mathbf{_\perp} = (Id - \bb_{eq} {\bb_{eq}}^T)\bB$ and $\bE\mathbf{_\perp} = (Id - \bb_{eq} {\bb_{eq}}^T) \bE$, respectively. The associated vector potentials of these magnetic fields are $\bA_{eq}$ and $\bA$, so that $\bB_{eq} = \nabla\times \bA_{eq}$ and $\bB = \nabla\times \bA$. We also denote the total magnetic field as $\bB_{tot} = \bB_{eq} + \bB$. The electrostatic potential is given by $\phi$ and the charge of the particle species by $q_e$, which is the electron charge in the present context. The particle kinetic energy for the drift-kinetic electrons is denoted by $K = K_0 + K_1 + K_2$. Here, $K_0$ is the component independent of $\bE$ and $\bB$, $K_1$ depends linearly on $\bE$ and $\bB$, and $K_2$ depends quadratically on $\bE$ and $\bB$. These are given by:
\begin{align}
K_0 &= \frac{1}{2}m_e v_{\parallel}^2 + \mu |\bB_{eq}|, \label{def:K0}\\
K_1 &= \mu \bb_{eq} \cdot \bB,  \label{def:K1}\\
K_2 &=  \left( \mu |\bB_{eq}| - m_e v_{\parallel}^2 \right) \frac{|\bB\mathbf{_\perp}|^2}{2|\bB_{eq}|^2} - \frac{m_e|\bE\mathbf{_\perp}|^2}{2|\bB_{eq}|^2} - \frac{m_e v_{\parallel} \bE \times \bb_{eq} \cdot \bB}{ \bB_{eq}^2}. \label{def:K2}
\end{align}
Here, $m_e$ is the particle mass of the electron species, $\bv_{e,gc}$ is the particle guiding center velocity and $v_{\parallel} = \bv_{e,gc}\cdot\bB_{eq}$ is its component that is parallel to the equilibrium magnetic field. $\mu$ is the magnetic moment of the particle. We also introduce the effective magnetic field $\bB^{\ast}$ and its associated vector potential $\bA^{\ast}$. These are given by:
\begin{align}
\bA^{\ast} &= \bA + \bA_{eq} + \frac{m_e}{q_e} v_{\parallel} \bb_{eq}, \\ 
\bB^{\ast} &= \nabla \times \bA = \bB + \bB_{eq} + \frac{m_e}{q_e} v_{\parallel} \nabla \times \bb_{eq}.
\end{align}
Here, the $\frac{m_e}{q_e} v_{\parallel} \nabla \times \bb_{eq}$ term is the field-line geometry correction, that arises from the removal of electron gyromotion in this model. Finally, the Lagrangian for the drift-kinetic model has been described by \citet{meng2025}. For the quasineutral model considered here, the electric field energy term must be removed and hence the Lagrangian for the drift-kinetic model in this case is given by:
\begin{multline}\label{eq:action}
	L_{e}(\bA(t), \phi(t), \bX_{e,gc}(t),\dot{\bX}_{e,gc}(t),V_{\parallel}(t), \mu; f_{e,gc,0}) \\
	= \int  q_e \left(\bA^{\ast}(t,\bX_{e,gc}(t),V_{\parallel}(t),\bA(t,\bX_{e,gc}(t))) \cdot \dot{\bX}_{e,gc}(t) - \phi(t,\bX_{e,gc}(t))\right) f_{e,gc,0} B^{\ast}_\parallel{_0} \dd \bx_{0} \dd v_{\parallel}{_0} \dd \mu \\
	- \int K_0(\bX_{e,gc}(t), V_{\parallel}(t),\mu) f_{e,gc,0} B^{\ast}_\parallel{_0} \dd \bx_{0} \dd v_{\parallel}{_0} \dd \mu \\
	- \int (K_1+K_2)(\bX_{e,gc}(t), V_{\parallel}(t),\bE(t, \bX_{e,gc}(t)), \bB(t,\bX_{e,gc}(t)),\mu) f_{e,gc,0} B^{\ast}_\parallel{_0} \dd \bx_{0} \dd v_{\parallel}{_0} \dd \mu.
	\end{multline}
 Here, $f_{e,gc}$ is the guiding center phase-space distribution function expressed using the guiding center volume element $B^{\ast}_\parallel q_e \dd \bx dv_{\parallel} \dd \mu$.
$B^{\ast}_\parallel = \bB^{\ast} \cdot \bb_{eq} $ is the component of $\bB^{\ast}$ that is parallel to $\bB_{eq}$. For the ions, we must use the Lagrangian for the quasineutral fully kinetic model, developed by \citet{Tronci2015Neutral-Vlasov-}. This is given by:
\begin{multline}\label{eq:action}
	L_{i}(\bA(t),\phi(t), \bX_i(t),\dot{\bX_i}(t),\bV_i(t); f_{i,0}) 
	= \\ \int \left((m_i \bV_i(t) + q_i\bA(t,\bX_i(t))) \cdot \dot{\bX_i}(t) - \frac 12 m_i V_i^2- q_i \phi(t,\bX_i(t))\right) f_{i,0} \dd \bx_{0} \dd \bv_{0}, 
\end{multline}
Here, $m_i$ and $q_i$ are the ion particle mass and charge, respectively, while $f_i$ is the phase-space distribution function for the ions. The Lagrangian accounting for the magnetic field energy is given by:
\begin{equation}
    L_{f}(\bA(t)) = - \int \left(\frac{1}{2\mu_0} \left| \nabla\times\bA_{eq}(\bx) + \nabla\times\bA(t,\bx)\right|^2 \right) \dd \bx.
\end{equation}
In the above equations, $\bB_{eq} = \nabla\times\bA_{eq}$ and $\bB = \nabla\times\bA$. Also, the electric field can be written in terms of the potentials as $\bE = - \fracp{\bA}{t} - \nabla\phi$. The total Lagrangian for the complete hybrid model is therefore given by:
\begin{equation}
    L = L_{e} + L_{i} + L_{f}.
\end{equation}
    
\subsection{Governing equations}\label{sec:goveqns}
We obtain the physical system of governing equations from the above action. Note that these equations can also be obtained from the non-quasineutral hybrid drift-kinetic model presented in the work by \citet{meng2025}, by simply taking the formal limit $\epsilon_0 \to 0$, where $\epsilon_0$ is the permittivity of vacuum. Taking variations with respect to $\bX$ and $\bV$ yields the Euler-Lagrange equations for the particles, $\frac{\partial}{\partial t} \frac{\partial L}{\partial\dot{\bX}} = \frac{\partial L}{\partial \bX}$ and $\frac{\partial L}{\partial \bV} = \bf 0$. Taking variations with respect to the electron guiding-center positions and velocities, $\bX_{e,gc}$ and $\bv_{e,gc}$, gives the following equations of motion for the drift-kinetic electrons:
\begin{align}
	\fract{\bX_{e,gc}}{t} = \bV_{e,gc} &=V_{\parallel}\frac{\bB^{\ast}}{B_{\parallel}^\ast} 
	+ \frac{1}{B_{\parallel}^\ast} \left(\bE\times\bb_{eq} -\frac {\mu}{q_e} \nabla B_{\parallel,tot} \times\bb_{eq} \right),  \label{eq:Xedot}\\
	\fract{V_{\parallel}}{t}&= \frac {q_e}{m_e}\frac{\bB^{\ast}}{B_{\parallel}^\ast} \cdot \left(\bE - \frac {\mu}{q_e} \nabla B_{\parallel,tot}\right) = a_{e,gc},\label{eq:Vedot}
\end{align}
where $B_{\parallel,tot} = \bB^{\ast} \cdot \bb_{eq}$ and $\bV_{e,gc}$ and $a_{e,gc}$ are the guiding center velocity and acceleration, respectively. Equations \eqref{eq:Xedot} and \eqref{eq:Vedot} are the characteristic equations of the drift-kinetic Vlasov equation for the electron guiding centers, given by:
\begin{equation}\label{eq:VlasovDKe}
    \fracp{f_{e,gc}}{t} + \bV_{e,gc}\cdot\nabla_x f_{e,gc} + a_{e,gc}\fracp{f_{e,gc}}{v_{\parallel}}=0.
\end{equation}
Similarly, taking variations with respect to the ion particle positions and velocities, $\bX_i$ and $\bV_i$, gives the following equations of motion for the ions:
\begin{align}
	\fract{\bX_i}{t}&=\bV_i, \label{eq:Xidot}\\
	\fract{\bV_i}{t}&= \frac {q_i}{m_i}\left(\bE + \bV_i\times \bB_{tot}\right) = a_i,\label{eq:Vidot}
\end{align}
where $a_i$ is the ion acceleration. The solutions of equations \eqref{eq:Xidot} and \eqref{eq:Vidot} are the characteristic equations of the fully kinetic Vlasov equation for the ions, given by:
\begin{equation}\label{eq:VlasovFKi}
    \fracp{f_i}{t} + \bV_i\cdot\nabla_x f_i + a_i\cdot\fracp{f_i}{\bv_i}=0.
\end{equation}
Here, $f_{e,gc}$ and $f_i$ are the velocity-space distribution functions for the electron guiding centers and ions, respectively. The variations in $\bA$ yield the quasineutral Amp\`ere equation:
\begin{equation}\label{eq:AmpereQN}
	\nabla\times\bB_{tot} = \mu_0\bJ = \mu_0(\bJ_i + \bJ_{e,gc}),
\end{equation}
where $\bJ$ is the total current density, and $\bJ_{e,gc}$ and $\bJ_i$ are the electron and ion current densities, respectively, given by:
\begin{align}
    \bJ_{e,gc} &= q_e \int f_{e,gc} \bv_{e,gc} B^{\ast}_\parallel \dd v_{\parallel} \dd \mu,\label{eq:Je}\\
	\bJ_i &= q_i \int f_i \bv \dd \bv\label{eq:Ji}.
\end{align}
The variations in $\phi$ yield the quasineutrality condition:
\begin{equation}\label{eq:QN-condition}
	\rho = \rho_i + \rho_{e,gc} = 0,
\end{equation}
where $\rho$ is the total charge density, and $\rho_{e,gc}$ and $\rho_i$ are the electron and ion charge densities, respectively, given by:
\begin{align}
    \rho_{e,gc} &= q_e \int f_{e,gc} B^{\ast}_\parallel \dd v_{\parallel} \dd \mu \label{eq:rhoe},\\
	\rho_i &= q_i \int f_i \dd\bv. \label{eq:rhoi}
\end{align}
Equation \eqref{eq:QN-condition} above
also implies the quasineutral continuity equation $\nabla \cdot (\bJ_i + \bJ_{e,gc}) = 0$ which can also be obtained by taking the dot product of equation \eqref{eq:AmpereQN}. Moreover, the Faraday's equation follows directly from the definition of the fields from the potentials and is given by:
\begin{equation}
\fracp{\bB}{t} + \nabla \times \bE = {\bf 0}, \label{eq:qnfaraday}
\end{equation}
since the temporal derivative of $\bB_{eq}$ is zero. Similarly, the Gauss law for magnetism follows from the definition of the magnetic field and is given by:
\begin{equation}
\nabla \cdot \bB = 0. \label{eq:qngaussB}
\end{equation}
Taking variations of the Lagrangian typically also generates extra polarization and magnetization terms. However, in the present work, these terms have been assumed to be negligible and are therefore not accounted for. For a more complete derivation and detailed treatment of these terms, the reader is referred to the work by \citet{meng2025}. In the present work, we also assume $\bB_{eq}$ to be uniform. Therefore, in this case, $\nabla \times \bB_{eq}$ in equation \eqref{eq:AmpereQN} and $\nabla \cdot \bB_{eq}$ in equation \eqref{eq:qngaussB} both simplify to zero.

\subsection{The electric field equations}\label{sec:efieldeqn}
Without the displacement current and electron polarization current terms in the Amp\`ere's equation, the electric field cannot be evolved using its time derivative. Other equations are therefore required to obtain the electric field at any time instant. We first write the electric field as a sum of its components perpendicular and parallel to $\bB_{eq}$ i.e. $\bE = \bE_\perp + \bE_\parallel$. Also $\bE_\parallel$ can be written as $\bE_\parallel = E_\parallel \bb_{eq}$, where $E_\parallel$ is the scalar value of $\bE_\parallel$.

\subsubsection{Equation for $\bE\mathbf{_\perp}$}\label{sec:eperpeqn}
The $\bE\mathbf{_\perp}$ equation can be directly obtained from the Amp\`ere equation, i.e. equation \eqref{eq:AmpereQN}. Using the $\bv_{e,gc}$ definition from equation \eqref{eq:Xedot}, and the fact that $\bE \times \bb_{eq} = \bE\mathbf{_\perp} \times \bb_{eq}$, the drift-kinetic electron current density defined in equation \eqref{eq:Je} can be rewritten as:
\begin{equation}
    \bJ_{e,gc} = q_e \int v_{\parallel}{\bB^{\ast}} f_{e,gc} \dd v_{\parallel} \dd \mu + 
    \int \left(q_e\bE\mathbf{_\perp} - \mu \nabla B_{\parallel,tot} \right)\times\bb_{eq} f_{e,gc} \dd v_{\parallel} \dd \mu\label{eq:Je_vgc}.
\end{equation}
We now take the cross product of $\bb_{eq}$ with equation \eqref{eq:Je_vgc}. Using the relations $\bb_{eq} \times (\bE\mathbf{_\perp} \times \bb_{eq}) = \bE\mathbf{_\perp}$ and $\bb_{eq} \times (\nabla B_{\parallel,tot} \times \bb_{eq}) = \nabla_\perp B_{\parallel,tot}$, we get:
\begin{equation}
     \int q_e\bE\mathbf{_\perp} f_{e,gc} \dd v_{\parallel} \dd \mu = \bb_{eq} \times\bJ_{e,gc} - q_e\int \left(v_{\parallel}{\bb_{eq} \times\bB^{\ast}} - \frac{\mu}{q_e} \nabla_\perp B_{\parallel,tot} \right) f_{e,gc} \dd v_{\parallel} \dd \mu\label{eq:bext_x_Je_vgc}.
\end{equation}
Finally, using equation \eqref{eq:AmpereQN} to substitute for $\bJ_{e,gc}$, we get the final equation to solve for $\bE\mathbf{_\perp}$:
\begin{equation}\label{eq:Eperp_eqn}
     \mu_0 \int q_e \bE\mathbf{_\perp} f_{e,gc} \dd v_{\parallel} \dd \mu = \bb_{eq} \times(\nabla\times\bB - \mu_0\bJ_i) - \mu_0q_e\int \left(v_{\parallel}{\bb_{eq} \times\bB^{\ast}} - \frac{\mu}{q_e} \nabla_\perp B_{\parallel,tot} \right) f_{e,gc} \dd v_{\parallel} \dd \mu.
\end{equation}
Obtaining $\bE\mathbf{_\perp}$ by solving equation \eqref{eq:Eperp_eqn} ensures the satisfaction of Amp\`ere's law given in equation \eqref{eq:AmpereQN} and hence the divergence-free nature of the current density.

\subsubsection{Equation for $E\mathbf{_\parallel}$}\label{sec:epareqn}
Taking the curl of the Faraday's equation, i.e. \eqref{eq:qnfaraday} and the time derivative of the Amp\`ere equation, i.e. equation \eqref{eq:AmpereQN}, and combining the two equations, we obtain:
\begin{equation} \label{eq:qncurlcurl_fullE}
    \nabla \times \nabla \times \bE = - \mu_0\left(\fracp{\bJ_i}{t} + \fracp{\bJ_{e,gc}}{t}\right).
\end{equation}
Taking the dot product of equation \eqref{eq:qncurlcurl_fullE} with $\bb_{eq}$, we get:
\begin{equation} \label{eq:qncurlcurl}
    \bb_{eq}\cdot (\nabla \times \nabla \times \bE) = \nabla_\parallel(\nabla_\perp \cdot \bE_\perp) - \nabla_\perp^2 E_\parallel = - \mu_0 \bb_{eq}\cdot\left(\fracp{\bJ_i}{t} + \fracp{\bJ_{e,gc}}{t}\right).
\end{equation}
Multiplying the ion Vlasov equation \eqref{eq:VlasovFKi} with $q_i \bv_i$ and integrating over the velocity space gives:
\begin{equation}\label{eq:qndjdt}
    \fracp{\bJ_i}{t} = \frac{q_i}{m_i} (\rho_i\bE + \bJ_i \times \bB - \nabla \cdot \mathbb{S}_i),
\end{equation}
where $\rho_i$ and $\bJ_i$ are the ion charge and ion current densities defined in equations \eqref{eq:rhoi} and \eqref{eq:Ji}, respectively. The term $\mathbb{S}_i$ is the contribution of the ion species to the stress tensor, given by:
\begin{equation}
    \mathbb{S}_i = m_i \int f_i \bv_i \otimes \bv_i d\bv. \label{eq:S_i}
\end{equation}
The term $\bb_{eq} \cdot \fracp{\bJ_{e,gc}}{t}$ becomes $\fracp{J_{e,gc\parallel}}{t}$, where $J_{e,gc\parallel}$ is given by:
\begin{equation}
    J_{e,gc\parallel} = q_e \int v_{\parallel} f_{e,gc} B^{\ast}_\parallel \dd v_{\parallel} \dd \mu.\label{eq:Je_par}\\
\end{equation}
To obtain $\fracp{J_{e,gc\parallel}}{t}$, we multiply the drift-kinetic electron Vlasov equation \eqref{eq:VlasovDKe} with $q_e v_{\parallel}$ and integrate over the velocity space to get:
\begin{equation}\label{eq:djpardt}
    \fracp{J_{e,gc\parallel}}{t} + \nabla\cdot \left(\boldsymbol{p}^\ast + (\eta\bE- Q\nabla B_{\parallel,tot})\times\bb_{eq}\right)
    - \boldsymbol{\alpha}^\ast \cdot \bE +\boldsymbol{\mu}^\ast\cdot\nabla B_{\parallel,tot}
    = 0,
\end{equation}
where we define:
\begin{align}
	\boldsymbol{p}^\ast(\bx) &= q_e \int v_{\parallel}^2 f_{e,gc} \bB^{\ast} \dd v_{\parallel}\dd\mu, \\
	\eta(\bx) &=   q_e \int v_{\parallel}  f_{e,gc} \dd v_{\parallel}\dd\mu , \\
    Q(\bx) &=  \int v_{\parallel}\mu f_{e,gc} \dd v_{\parallel}\dd\mu, \\
    \boldsymbol{\alpha}^\ast(\bx) &= \frac{q_e^2}{m_e} \int f_{e,gc} \bB^{\ast} \dd v_{\parallel} \dd\mu \label{eq:djpt}, \\
    \boldsymbol{\mu}^\ast(\bx) &= \frac{q_e}{m_e} \int \mu f_{e,gc} \bB^{\ast} \dd v_{\parallel} \dd\mu.
\end{align}
In equation \eqref{eq:djpt}, the term $\boldsymbol{\alpha}^\ast \cdot \bE$ reduces to $(\boldsymbol{\alpha}^\ast \cdot \bE_\perp + \boldsymbol{\alpha}^\ast \cdot\bb_{eq} E_\parallel)$. The term $\boldsymbol{\alpha}^\ast \cdot\bb_{eq}$ becomes:
\begin{equation}\label{eq:alpha_dot_bext}
    \boldsymbol{\alpha}^\ast \cdot\bb_{eq} = \frac{q_e^2}{m_e} \int f_{e,gc} B^{\ast}_\parallel \dd v_{\parallel} \dd\mu = \frac{q_e}{m_e} \rho_{e,gc}.
\end{equation}
Therefore, substituting \eqref{eq:qndjdt}, \eqref{eq:djpardt} and \eqref{eq:alpha_dot_bext} in \eqref{eq:qncurlcurl}, the equation for $E_\parallel$ becomes:
\begin{multline} \label{eq:Epar_eqn}
    - \nabla_\perp^2 E_\parallel + \mu_0 \left(\frac{q_i^2}{m_i}n_i + \frac{q_e^2}{m_e}n_{e,gc}\right)E_\parallel = \mu_0 \bb_{eq}\cdot\left(\frac{q_i}{m_i} (-\bJ_i \times \bB + \nabla \cdot \mathbb{S}_i)\right) +\\ \mu_0 \left(\nabla\cdot \left(\boldsymbol{p}^\ast + (\eta\bE_\perp- Q\nabla B_{\parallel,tot})\times\bb_{eq}\right)
    - \boldsymbol{\alpha}^\ast \cdot \bE_\perp +\boldsymbol{\mu}^\ast\cdot\nabla B_{\parallel,tot}\right) - \nabla_\parallel(\nabla_\perp \cdot \bE_\perp).
\end{multline}
Here, $n_i = q_i \rho_i$ and $n_{e,gc} = q_e \rho_{e,gc}$ are the ion and electron number densities, respectively. Knowing $\bE_\perp$ from equation \eqref{eq:Eperp_eqn}, we can solve for $E_\parallel$ using equation \eqref{eq:Epar_eqn}. The left hand side of equation \eqref{eq:Epar_eqn} gives a symmetric, positive definite matrix after discretization, making it a determinate and non-singular system with a unique solution. The magnetic moment, $\mu$, is assumed to be zero in the work presented here. This simplifies $Q(\bx)$ and $\boldsymbol{\mu}^\ast(\bx)$ in equation \eqref{eq:Epar_eqn} and the $\nabla B_{\parallel,tot}$ terms in equations \eqref{eq:Xedot}, \eqref{eq:Vedot} and \eqref{eq:Eperp_eqn} to zero. Going forward, this term and its discretized equivalent will not be shown in any of the equations in the upcoming sections.

%%%%%%%%%%%%%%%%%%%%%%%%%%%%%%%%%%%%%%%%%%%%%%%%%%%%%%%%%

%%%%%%%%%%%%%%%%%%%%%%%%%%%%%%%%%%%%%%%%%%%%%%%%%%%%%%%%%
\section{Structure-preserving spatial discretization of fields}\label{sec:fielddiscretization}
\begin{figure}
    \centering
    \includegraphics[width=0.95\linewidth]{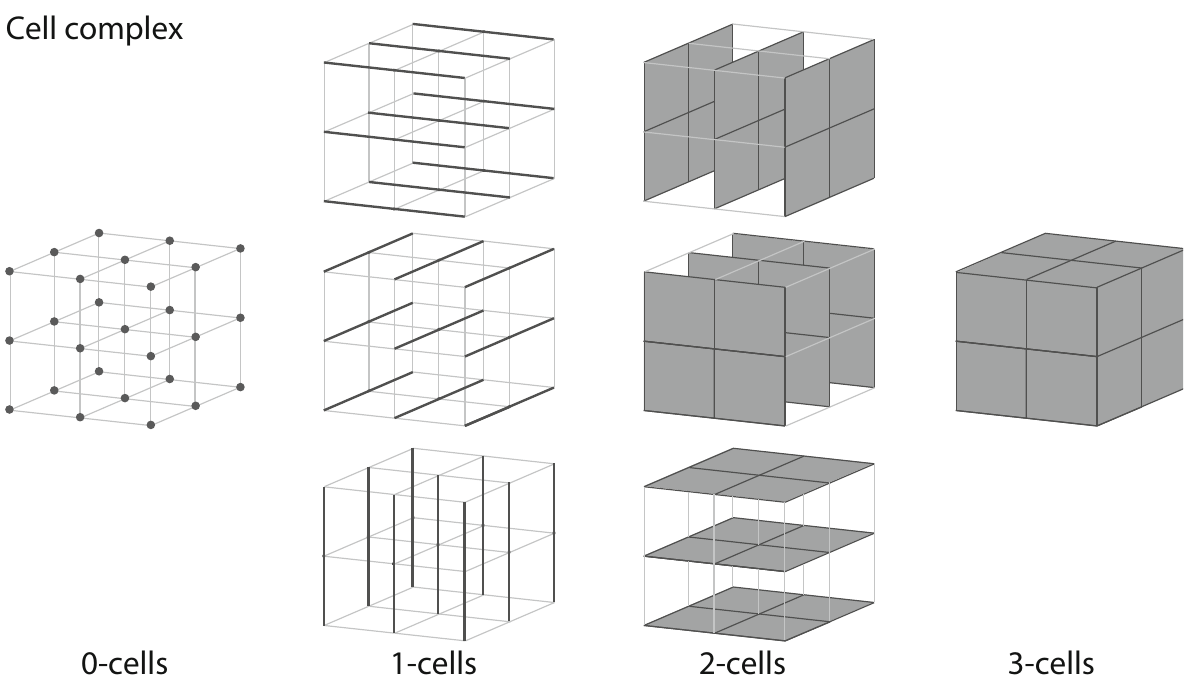}
    \caption{Spaces for defining discrete variables on a primal Cartesian grid.}
    \label{fig:CellComplexes}
\end{figure}

We use a dual grid approach for the spatial discretization of our model, wherein the dual or adjoint grid vertices are the barycenters of the primal grid cells. The numerical values for the field variables are located on spaces based on points, edges, faces and volumes of hexahedral (3D) or quadrilateral (2D) cells. This discretization, based on mimetic finite differences, has already been used in \texttt{GEMPICX} for discretizing the fully kinetic \citep{kormann2024}, drift-kinetic \citep{meng2025} and fully kinetic quasineutral \citep{narechania2026FKQN} Vlasov-Maxwell models. The nodal, edge, face and volume spaces on the primal grids are denoted by $\cC_0$, $\cC_1$, $\cC_2$ and $\cC_3$, respectively. Similarly, those on the dual grids are denoted by $\tilde \cC_0$, $\tilde \cC_1$, $\tilde \cC_2$ and $\tilde \cC_3$, respectively. A duality exists between the primal and duality spaces. The nodes, edges, faces and cell volumes on the primal grid are each uniquely associated with the cell volumes, faces, edges and nodes on the dual grid, respectively. On the primal grid, the electric field $\bE$ and the vector potential $\bA$ are defined as edge-integrals and the magnetic field $\bB$ is defined as face-integrals. On the dual grid, the magnetic field intensity $\bH$ is defined as edge-integrals, the current density $\bJ$ and electric field intensity $\bD$ are defined as face-integrals, and the charge density $\rho$ is defined as volume-integrals. Such an approach allows for exact preservation of certain discretized quantities. For example, the discrete divergence of $\bJ$ becomes a cell-integral on the dual grid, which is the same space where the charge density $\rho$ is defined. These spaces are shown in Figure \ref{fig:CellComplexes} for the primal grid.

\subsection{Reduction operators}\label{sec:reduction}
The reduction operators, also called restriction operators, relate a continuous field to its discretized counterparts defined on nodes, edges, faces and volumes. On the primal grid:
\begin{itemize}
	\item $\cR_0$ associates a scalar field to its values at primal grid nodes:
    \begin{align}
    \cR_0(\psi)_{i,j,k} = \psi(x_i, y_j, z_k) = {\bolduppsi}_{i,j,k}.
    \end{align}
    Here, $\bolduppsi \in \cC_0$.

    \item $\cR_1$ associates a vector field to its integrals on primal grid edges. For the $x-$edge, we have:
    \begin{gather}
    \cR_1^x(E_x)_{i,j,k} = \int_{x_i}^{x_{i+1}} E_x(x, y_j, z_k) \dd x = {\arr{E}}_{i+1/2,j,k} = {\arr{E}}^x_{i,j,k}.
    \end{gather}
    Similarly, $\cR_1^y(E_y)_{i,j,k}$ and $\cR_1^z(E_z)_{i,j,k}$ can be defined. The complete operator becomes: \\$\cR_1(\bE) = (\cR_1^x(E_x), \cR_1^y(E_y), \cR_1^z(E_z)) = ({\arr{E}}^x,{\arr{E}}^y,{\arr{E}}^z) = \arr{E} \in \cC_1$.

    \item $\cR_2$ associates a vector field to its integrals on primal grid faces. For the $x-$edge, we have:
    \begin{gather}
    \cR_2^x(B_x)_{i,j,k} = \int_{z_k}^{z_{k+1}} \int_{y_j}^{y_{j+1}} B_x(x_i, y, z) \dd y\dd z = {\arr{B}}_{i,j+1/2,k+1/2} = {\arr{B}}^x_{i,j,k}.
    \end{gather}
    Similarly, $\cR_2^y(B_y)_{i,j,k}$ and $\cR_2^z(B_z)_{i,j,k}$ can be defined. The complete operator becomes: \\$\cR_2(\bB) = (\cR_2^x(B_x), \cR_2^y(B_y), \cR_2^z(B_z)) = ({\arr{B}}^x,{\arr{B}}^y,{\arr{B}}^z) = \arr{B} \in \cC_2$.

	\item $\cR_3$ associates a scalar field to its volume integrals over primal grid cells:
    \begin{gather}
    \cR_3(\rho)_{i,j,k} = \int_{z_k}^{z_{k+1}} \int_{y_j}^{y_{j+1}} \int_{x_i}^{x_{i+1}} \rho(x, y, z) \dd x\dd y\dd z = {\bm{\uprho}}_{i,j,k}.
    \end{gather}
    Here, $\bm{\uprho} \in \cC_3$.
\end{itemize}
Similarly, reduction/restriction operators are also defined on the dual grid as follows:
\begin{itemize}
	\item $\tilde\cR_0$ associates a scalar field to its values at dual grid nodes:
    \begin{gather}
    \tilde\cR_0(\phi)_{i,j,k} = \phi(x_{i+1/2}, y_{j+1/2}, z_{k+1/2}) = \tilde{\boldupphi}_{i+1/2,j+1/2,k+1/2}.
    \end{gather}
    Here, $\tilde{\boldupphi} \in \tilde\cC_0$.

    \item $\tilde\cR_1$ associates a vector field to its integrals on dual grid edges. For the $x-$edge, we have:
    \begin{gather}
    \tilde\cR_1^x(H_x)_{i,j,k} = \int_{x_{i-1/2}}^{x_{i+1/2}} H_x(x, y_{j+1/2}, z_{k+1/2}) \dd x = \tilde{\arr{H}}_{i,j+1/2,k+1/2} = \tilde{\arr{H}}^x_{i,j,k}.
    \end{gather}
    Similarly, $\tilde\cR_1^y(H_y)_{i,j,k}$ and $\tilde\cR_1^z(H_z)_{i,j,k}$ can be defined. The complete operator becomes: \\$\tilde\cR_1(\bH) = (\tilde\cR_1^x(H_x), \tilde\cR_1^y(H_y), \tilde\cR_1^z(H_z)) = (\tilde{\arr{H}}^x, \tilde{\arr{H}}^y, \tilde{\arr{H}}^z) = \tilde{\arr{H}}  \in \tilde\cC_1$.

    \item $\tilde\cR_2$ relates a vector field to its integrals on dual grid faces. For the $x-$edge, we have:
    \begin{gather}
    \tilde\cR_2^x(D_x)_{i,j,k} = \int_{z_{k-1/2}}^{z_{k+1/2}} \int_{y_{j-1/2}}^{y_{j+1/2}} D_x(x_{i+1/2}, y, z) \dd y\dd z = \tilde{\arr{D}}_{i+1/2,j,k} = \tilde{\arr{D}}^x_{i,j,k}.
    \end{gather}
    Similarly $\tilde\cR_2^y(D_y)_{i,j,k}$ and $\tilde\cR_2^z(D_z)_{i,j,k}$ can be defined. The complete operator becomes: \\$\tilde\cR_2(\bD) = (\tilde\cR_2^x(D_x), \tilde\cR_2^y(D_y), \tilde\cR_2^z(D_z)) = (\tilde{\arr{D}}^x, \tilde{\arr{D}}^y, \tilde{\arr{D}}^z) = \tilde{\arr{D}} \in \tilde\cC_2$.

	\item $\tilde\cR_3$ relates a scalar field to its volume integrals over dual grid cells:
    \begin{gather}
    \tilde\cR_3(\rho)_{i,j,k} = \int_{z_{k-1/2}}^{z_{k+1/2}} \int_{y_{j-1/2}}^{y_{j+1/2}} \int_{x_{i-1/2}}^{x_{i+1/2}} \rho(x, y, z) \dd x\dd y\dd z = \tilde{\bm{\uprho}}_{i,j,k}.
    \end{gather}
    Here, $\tilde{\bm{\uprho}} \in \tilde\cC_3$.
\end{itemize}

\subsection{Hodge operators}\label{sec:hodge}
Field variables can be projected from primal grid spaces to their dual counterparts or vice versa using Hodge operators. Hodge operators of an arbitrary order of accuracy can be obtained using the method prescribed by \citet{kormann2024}. In the current work, only second-order Hodge projections have been used and this simply constitutes the use of scaling factors to relate variables between primal and dual spaces. Consider arbitrary variables $\bK$, $\bL$, $\bM$ and $\bN$, defined on primal nodes, edges, faces and cell volumes, respectively. Their corresponding dual mappings, defined on dual cell volumes, faces, edges and nodes are denoted as $\tilde\bK$, $\tilde\bL$, $\tilde\bM$ and $\tilde\bN$, respectively. These would be given by:
\begin{gather}
  \tilde{\arr{K}} = \Delta x\, \Delta y\, \Delta z \arr{K}, \notag \\
  \tilde{\arr{L}}^x = \frac{\Delta y\, \Delta z}{\Delta x} \arr{L}^x, \quad
  \tilde{\arr{L}}^y = \frac{\Delta x\, \Delta z}{\Delta y} \arr{L}^y, \quad
  \tilde{\arr{L}}^z = \frac{\Delta x\, \Delta y}{\Delta z} \arr{L}^z, \notag \\
  \tilde{\arr{M}}^x = \frac{\Delta x}{\Delta y\, \Delta z} \arr{M}^x, \quad
  \tilde{\arr{M}}^y = \frac{\Delta y}{\Delta x\, \Delta z} \arr{M}^y, \quad
  \tilde{\arr{M}}^z = \frac{\Delta z}{\Delta x\, \Delta y} \arr{M}^z, \notag \\
  \tilde{\arr{N}} = \frac{1}{\Delta x\, \Delta y\, \Delta z} \arr{N}.
\end{gather}
We can therefore define Hodge operators $\matH_0$, $\matH_1$, $\matH_2$, and $\matH_3$, and their corresponding inverses $\tilde \matH_3$, $\tilde \matH_2$, $\tilde \matH_1$, and $\tilde \matH_0$, respectively:
\begin{gather}
  \tilde{\arr{K}} = \matH_0 \arr{K},\;\hspace{0.5cm} \arr{K} = \tilde \matH_3 \tilde{\arr{K}}, \notag \\
  \tilde{\arr{L}} = \matH_1 \arr{L},\;\hspace{0.5cm} \arr{L} = \tilde \matH_2 \tilde{\arr{L}}, \notag \\
  \tilde{\arr{M}} = \matH_2 \arr{M},\;\hspace{0.5cm} \arr{M} = \tilde \matH_1 \tilde{\arr{M}}, \notag \\
  \tilde{\arr{N}} = \matH_3 \arr{N},\;\hspace{0.5cm} \arr{N} = \tilde \matH_0 \tilde{\arr{N}}.
\end{gather}
Accordingly, the electric and magnetic fields can be related to their intensities as follows:
\begin{gather}\label{eq:hodge_relns}
  \tilde{\arr{D}} = \matH_1 \arr{E},\;\hspace{0.5cm} \arr{E} = \tilde \matH_2 \tilde{\arr{D}}, \notag \\
  \tilde{\arr{H}} = \matH_2 \arr{B},\;\hspace{0.5cm} \arr{B} = \tilde \matH_1 \tilde{\arr{H}}.
\end{gather}
These various Hodge operators are simply the appropriate scaling factors multiplied by the identity matrix. This corresponds to the classical Yee scheme. While we use only the second-order scheme here, the reader is referred to the work by \citet{kormann2024} for details of higher-order Hodge operators and their implementation.

\subsection{Discrete scalar products}\label{sec:scalarproducts}
Discrete approximations of $L^2$ inner products of scalar or vector fields can be calculated using discrete scalar products between primal field variables and their dual grid Hodge projections. For the variables $\bK$, $\bL$, $\bM$ and $\bN$ defined above, these discrete scalar products are as follows:
\begin{equation}
  \arr{K} \cdot \tilde{\arr{K}} = \sum_{i,j,k} \arr{K}_{i,j,k} \tilde{\arr{K}}_{i,j,k},
\end{equation}
\begin{gather}
  \arr{L} \cdot \tilde{\arr{L}} 
  = \arr{L}^x \cdot \tilde{\arr{L}}^x 
   + \arr{L}^y \cdot \tilde{\arr{L}}^y 
   + \arr{L}^z \cdot \tilde{\arr{L}}^z \notag \\
  = \sum_{i,j,k} \big(
     \arr{L}_{i+1/2,j,k} \tilde{\arr{L}}_{i+1/2,j,k}
   + \arr{L}_{i,j+1/2,k} \tilde{\arr{L}}_{i,j+1/2,k}
   + \arr{L}_{i,j,k+1/2} \tilde{\arr{L}}_{i,j,k+1/2}
   \big),
\end{gather}
\begin{gather}
  \arr{M} \cdot \tilde{\arr{M}} 
  = \arr{M}^x \cdot \tilde{\arr{M}}^x 
   + \arr{M}^y \cdot \tilde{\arr{M}}^y 
   + \arr{M}^z \cdot \tilde{\arr{M}}^z \notag \\
  = \sum_{i,j,k} \big(
     \arr{M}_{i,j+1/2,k+1/2} \tilde{\arr{M}}_{i,j+1/2,k+1/2}
   + \arr{M}_{i+1/2,j,k+1/2} \tilde{\arr{M}}_{i+1/2,j,k+1/2} 
  %\hphantom{=
  %\sum_{i,j,k} \big(%}
   + \arr{M}_{i+1/2,j+1/2,k} \tilde{\arr{M}}_{i+1/2,j+1/2,k}
   \big),
\end{gather}

\begin{equation}
  \arr{N} \cdot \tilde{\arr{N}} = \sum_{i,j,k} \arr{N}_{i+1/2,j+1/2,k+1/2} \tilde{\arr{N}}_{i+1/2,j+1/2,k+1/2}.
\end{equation}

\subsection{Discrete gradient, curl and divergence}\label{sec:discretederivatives}
In order to define the discrete gradient, curl and divergence operators, we first define a one-dimensional discrete derivative or difference operator:
\begin{equation*}
	\mathbb{d}_{M_1} = \begin{pmatrix}
   -1 & 1 & 0 & \ldots & 0 \\
   0 & -1 & 1 & 0  & \\
   \vdots & & \ddots & \ddots & \\
   0 & & & -1 & 1 \\
   1 & 0 & \ldots & 0 & -1
   \end{pmatrix} \in \mathbb{R}^{M_1 \times M_1}.
\end{equation*}
We can represent a $\mathbb{d}$ matrix and an identity matrix of size $N$ as $\mathbb{d}_N$ and $\matI_N$, respectively. Consider a grid with $N_1$, $N_2$, $N_3$ points in the $x-$, $y-$ and $z-$ directions, respectively, such that $N = N_1 \times N_2 \times N_3$. For such a grid, the discrete gradient, curl, and divergence operators on the primal grid can be denoted by $\mathbb{G}$, $\mathbb{C}$ and $\mathbb{D}$. These are built using Kronecker products of the $\mathbb{d}$ and $\matI$ matrices of appropriate sizes as follows:

\begin{equation} \label{hD0}
\matG = \begin{pmatrix}
\matd_{N_1} \otimes \matI_{N_2} \otimes \matI_{N_3} \\
\matI_{N_1} \otimes \matd_{N_2} \otimes \matI_{N_3} \\
\matI_{N_1} \otimes \matI_{N_2} \otimes \matd_{N_3}
\end{pmatrix},
\end{equation}

\begin{equation}\label{hD12}
\matC = \begin{pmatrix}
\matO_{N} & - \matI_{N_1} \otimes \matI_{N_2} \otimes \matd_{N_3}  & \matI_{N_1} \otimes \matd_{N_2} \otimes \matI_{N_3}\\
\matI_{N_1} \otimes \matI_{N_2} \otimes \matd_{N_3}  & \matO_{N} &  -\matd_{N_1} \otimes \matI_{N_2} \otimes \matI_{N_3}\\
-\matI_{N_1} \otimes \matd_{N_2} \otimes \matI_{N_3}  & \matd_{N_1} \otimes \matI_{N_2} \otimes \matI_{N_3} & \matO_{N} \\
\end{pmatrix},
\end{equation}

\begin{equation} \label{hD3}
\matD = \begin{pmatrix}
\matd_{N_1} \otimes \matI_{N_2} \otimes \matI_{N_3} & \matI_{N_1} \otimes \matd_{N_2} \otimes \matI_{N_3} &  \matI_{N_1} \otimes \matI_{N_2} \otimes \matd_{N_3} \\
\end{pmatrix}.
\end{equation}
The adjoint operators of these operators on the dual grid are given by:
\begin{gather}
\tilde \matG = - \matD^T, \\
\tilde \matC = \matC^T, \\
\tilde \matD = - \matG^T.
\end{gather}
Due to our degrees of freedom being defined on discrete spaces associated with the de Rham complex, the discrete gradient, curl and divergence operators and their dual operators are always exact.
%%%%%%%%%%%%%%%%%%%%%%%%%%%%%%%%%%%%%%%%%%%%%%%%%%%%%%%%%

%%%%%%%%%%%%%%%%%%%%%%%%%%%%%%%%%%%%%%%%%%%%%%%%%%%%%%%%%
\section{Particle-in-Cell (PIC) discretization}\label{sec:particlediscretization}

Let us now consider the particle discretization of the Vlasov equation, where we write:
\begin{equation} \label{eq:fpart}
	f_s(t,\bx,\bv) = \sum_{p=1}^{N_s} w_p \delta(\bx -\bx_p(t))\delta(\bv -\bv_p(t)),
\end{equation}
where $w_p$ is the particle weight, $N_s$ is the number of particles of species `$s$' and $\delta$ is the Dirac delta function. This function is clearly not continuous and smooth. However, in order to use particle information to calculate fields such as $\rho$, $\bJ$ and $\mathbb{S}$ in their discretized forms, we would require a smoothing kernel, typically a spline. Such a spline would effectively spread out the influence of the particle's mass and charge over a volume centered at the particle location, rather than keep it concentrated at a single point. Hence, for a generic spline $S$, the smoothed particle charge density $\rho_s$ and particle current density $\bJ_s$, where the species `$s$' can be either ions or electrons, would be given by:
\begin{align} \label{eq:rhospart}
	\rho_s(t,\bx) &= \sum_{p=1}^{N_s} q_s w_{s,p} S(\bx -\bx_p(t)), \\
    \bJ_s(t,\bx) &= \sum_{p=1}^{N_s} q_s \bv_{s,p} w_{s,p} S(\bx -\bx_p(t)).
\end{align}
These splines would also be used to calculate electric and magnetic fields at particle locations using the discretized forms of the fields. The particles then obey equations \eqref{eq:Xedot} and \eqref{eq:Vedot}, or equations \eqref{eq:Xidot} and \eqref{eq:Vidot}, depending on the species. The splines used in this work are based on the well-known cardinal B-splines. Fundamental cardinal B-splines of an arbitrary degree $d$ centered at $x = 0$ can be defined recursively using the convolution:
\begin{equation}
S^{(d)}(x) = S^{(0)} * S^{(d-1)}(x) = \int_{-1/2}^{1/2} S^{(d-1)}(x - x') \, dx',
\end{equation}
with
\begin{equation}
S^{(0)}(x) =
\begin{cases}
1 & \text{if } -\frac{1}{2} \le x \le \frac{1}{2} \\
0 & \text{otherwise}
\end{cases}.
\end{equation}
Cardinal B-splines have the property:
\begin{equation}
\int_{-\infty}^{\infty} S^{(d)}(x) \, dx = 1.
\end{equation}
The coupling of particles and fields involves two kinds of B-splines: node splines and cell splines. These splines are built using the cardinal B-spline described above.

These splines are essential to evaluation of $\bE$, $\bB$ fields at particle positions and evaluation of particle-integrated field variables like charge density $\rho$ and current density $\bJ$, as defined in their respective spaces. Consider the mesh size in the $x-$direction to be denoted as $h$. Node splines are centered on grid nodes, and are defined as:
\begin{equation}
S_i^{n}(x_p) = S^d((x_p - x_i)/h).
\end{equation}
This spline is supported on the interval $[x_i - (d+1)h/2, x_i + (d+1)h/2]$, and its values are independent of the grid size.
Cell splines are centered on cell midpoints, and are defined as:
\begin{equation}
S_i^{c}(x_p) = \frac{1}{h}S^{d-1}\left(\frac{x_p - x_{i+1/2}}{h}\right).
\end{equation}
This spline is supported on the interval $[x_{i+1/2} - dh/2, x_{i+1/2} + dh/2]$, and its cell-integrals are independent of the grid size. Node and cell splines for the $y-$ and $z-$ directions can be similarly defined given the respective mesh sizes in those directions.

The particle charge densities $\rho_s$ are defined as cell volume integrals on the dual grid i.e. $\tilde{\bm{\uprho}}_s = \tilde\cR_3(\rho_s)  \in \tilde\cC_3$. These are calculated as:
\begin{equation}
\tilde{\bm{\uprho}}_s(t)_{i,j,k} = \tilde\cR_3(\rho_s(t))_{i,j,k} = \sum_{p=1}^{N_s} q_s w_{s,p} S_i^{n}(x_{s,p}(t)) S_j^{n}(y_{s,p}(t)) S_k^{n}(z_{s,p}(t)).
\end{equation}
The particle current densities $\bJ_s$ are defined as face integrals on the dual grid i.e. $\tilde{\arr{J}}_s = \tilde\cR_2(\bJ_s)  \in \tilde\cC_2$. The $x-$component, for instance, is calculated as:
\begin{equation}
\tilde{\arr{J}}_s^x(t)_{i+1/2,j,k} = \tilde\cR_2(J_s^x(t))_{i+1/2,j,k} = \sum_{p=1}^{N_s} q_s w_{s,p} v_{s,p,x} S_i^{c}(x_{s,p}(t)) S_j^{n}(y_{s,p}(t)) S_k^{n}(z_{s,p}(t)),\label{eq:depositjx}
\end{equation}
where $v_{s,p,x}$ is the $x-$component of the velocity of the kinetic ions or the drift-kinetic electrons. Similarly, the $y-$ and $z-$ components, i.e. $\tilde{\arr{J}}_s^y$ and $\tilde{\arr{J}}_s^z$ are calculated using the appropriate combinations of the cell and node splines. For the drift-kinetic electrons, the $\tilde{\arr{J}}_{e,gc}$ would also require calculation of electric and magnetic fields at the particle positions, which is described below. The terms $\bJ_i \times \bB$, $p^\ast$, $\eta \bE$, $\boldsymbol{\alpha}^\ast \cdot \bE$ and $\mathbb{S}_i$ are all similarly defined as face integrals on the dual grid. Gradients of these terms, wherever required, are simply obtained by replacing the appropriate $S^n$ or $S^c$ splines in these integrals with their respective spatial derivatives.

The smoothed electric and magnetic fields, $\bE^S = (E_x^S, E_y^S, E_z^S)$ and $\bB^S = (B_x^S, B_y^S, B_z^S)$, used to update particle velocities, are also calculated at particle positions using node and cell splines. These fields are estimated as spline-weighted linear combinations of the edge-integrated or face-integrated values of these fields. For example, the $x-$components of these fields at particle position, for species `$s$', are calculated as:
\begin{align}
E_x^S(\bx_{s,p}) &= \sum_{i,j,k} {\arr{E}}_{i+1/2,j,k} S_i^{c}(x_{s,p}) S_j^{n}(y_{s,p}) S_k^{n}(z_{s,p}), \\
B_x^S(\bx_{s,p}) &= \sum_{i,j,k} {\arr{B}}_{i,j+1/2,k+1/2} S_i^{n}(x_{s,p}) S_j^{c}(y_{s,p}) S_k^{c}(z_{s,p}).
\end{align}
Similarly, the $y-$ and $z-$ components are calculated using the appropriate combinations of node and cell splines.

%%%%%%%%%%%%%%%%%%%%%%%%%%%%%%%%%%%%%%%%%%%%%%%%%%%%%%%%%

%%%%%%%%%%%%%%%%%%%%%%%%%%%%%%%%%%%%%%%%%%%%%%%%%%%%%%%%%
\section{Semi-discretized governing equations}\label{sec:discreteqnvmmodel}
We now propose a semi-discrete action principle and take its variations to obtain the semi-discrete governing equations. The discretization is only in space, while all quantities are still varying continuously with time. The semi-discrete electric field equations are also derived.
\subsection{Semi-discrete action principle}\label{sec:discreteaction}
We use $\bX_i$ and $\bV_i$ to respectively denote positions and velocities of particles representing the fully kinetic ions. Similarly, for the drift-kinetic electrons, we use $\bX_{e,gc}$ and $\bV_{e,gc}$ for the guiding centers.
We follow the steps from the works by \citet{kormann2024} and \citet{meng2025} to derive the discrete action principle for the hybrid quasineutral model. The semi-discrete Lagrangians for the drift-kinetic electrons, fully kinetic ions, and the fields are given by:

\begin{multline}\label{eq:dk}
	\mathcal{L}_{h,e} = \sum_{p=1}^{{N}_e}
	  \left[w_{e,p} (m_{e} V_{\parallel,p} \bb_{eq}(\bX_{e,p}) + q_e (\arr{A}_{eq} + \arr{A})) \cdot \tilde{\mathcal{R}}_2  \left(\dot{\bX}_{e,p} S_p(\bX - \bX_{e,p})\right) \right. \\
	  \left. - q_e\arr{\boldupphi} \cdot \tilde{\mathcal{R}}_3(S_p(\bX - \bX_{e,p})) - \frac{m_e}{2} V_{\parallel,p}^2\right],
\end{multline}

\begin{gather}
	\mathcal{L}_{h,i} = \sum_{p=1}^{N_i} \left[w_{i,p} (m_i \bV_{i,p} + q_i (\arr{A}_{eq}+\arr{A})) \cdot \tilde{\mathcal{R}}_2  \left( \dot{\bX}_{i,p} S(\bX-\bX_{i,p})\right) - \frac 12 m_i V_{i,p}^2 - q_i \arr{\boldupphi} \cdot \tilde{\mathcal{R}}_3(S(\bX-\bX_{i,p})) \right], \label{eq:discLions}
\end{gather}

\begin{gather}
	\mathcal{L}_{h,f} = - \frac{1}{2\mu_0} (\mathbb{C}\arrA) \cdot (\mathbb{H}_1\mathbb{C}\arrA). \label{eq:discLfields}
\end{gather}
In the above equations, $N_i$ and $N_e$ are the number of ion and electron particles, respectively. The total Lagrangian is therefore $\mathcal{L}_h = \mathcal{L}_{h,e} + \mathcal{L}_{h,i} + \mathcal{L}_{h,f}$. We first take the variations with respect to $\bX_{e,gc,p}$ and $V_{\parallel,p}$ to obtain the discretized Euler-Lagrange equations for the drift-kinetic electrons. After rearranging them, we get the following equations of motion for the drift-kinetic electrons:
\begin{align}
	\fract{\bX_{e,gc,p}}{t} &= {V}_{\parallel,p} \frac{\bB^{\ast}(\bX_{e,gc,p})}{B_\parallel^{\ast}(\bX_{e,gc,p})}  + \frac{\bE_\perp^S(\bX_{e,gc,p})\times \bb_{eq} }{B_\parallel^{\ast}(\bX_{e,gc,p})} = \bV_{e,gc,p}, \label{eq:Xdotgcdisc} \\
	\fract{{V}_{\parallel,p}}{t} &= \frac{\bB^{\ast}(\bX_{e,gc,p})}{B_\parallel^{\ast}(\bX_{e,gc,p})}  \cdot \left(\frac{q_e}{m_e} \bE^S(\bX_{e,gc,p})\right) = \frac{q_e}{m_e} E_\parallel^S(\bX_{e,gc,p}).\label{eq:Vdotgcdisc}
\end{align}
The discretized Euler-Lagrange equations for the ion particles, obtained by taking the variations with respect to $\bX_{i,p}$ and $\bV_{i,p}$ are given by:
\begin{align}
	\fract{\bX_{i,p}}{t} &=\bV_{i,p}, \label{eq:Xdotidisc}\\
	\fract{\bV_{i,p}}{t} &= \frac {q_i}{m_i}\left(\bE^S(t,\bX_{i,p}) + \bV_{i,p} \times \bB^S(t,\bX_{i,p})\right).\label{eq:Vdotidisc}
\end{align}
In our discrete setting, the electromagnetic fields and the potentials, all defined on the primal grid, are related by:
\begin{equation}
    \arrE = -\frac{\dd \arrA}{\dd t} - \mathbb{G}\arr{\boldupphi}, ~~~~ \arrB = \mathbb{C}\arrA.
\end{equation}
This immediately yields the discrete Faraday equation:
\begin{equation} \label{eq:faraday_disc}
    \fract{\arrB}{t} + \mathbb{C}\arrE = 0.
\end{equation}
Similarly, the definition of the magnetic field, coupled with the discretization of the magnetic field on primal grid faces, allows us to write the discretized Gauss law of magnetism:
\begin{equation} \label{eq:gaussB_disc}
\mathbb{D}\arrB = 0.
\end{equation}
Taking the variations of $\mathcal{L}_{h}$ with respect to $\arrA$, we find Amp\`ere's law:
\begin{equation} \label{eq:ampere_disc}
    \mathbb{C}^\top\mathbb{H}_2\arrB = \mathbb{C}^\top\mathbb{H}_2 \mathbb{C}\arrA = \mu_0(\tilde{\arr{J}}_i + \tilde{\arr{J}}_{e,gc}),
\end{equation}
where the discretized current densities are given by:
\begin{align}
    \tilde{\arr{J}}_i(t) &= \sum_{p=1}^{N_i} w_{i,p} q_i \tilde{\mathcal{R}}_2  \left(\bV_{i,p}(t) S(\bX-\bX_{i,p}(t))\right), \label{eq:Ji_disc}\\
    \tilde{\arr{J}}_{e,gc}(t) &= \sum_{p=1}^{N_e} w_{e,p} q_e \tilde{\mathcal{R}}_2 \left(\bV_{e,gc,p}(t) S(\bX-\bX_{e,gc,p}(t))\right). \label{eq:Jgc_disc}
\end{align}
Taking the variations with respect to $\arr{\boldupphi}$ leads to the discretized quasineutrality condition:
\begin{equation} \label{eq:quasineutrality_disc}
    \tilde{\bm{\uprho}}_i(t) + \tilde{\bm{\uprho}}_{e,gc}(t) = 0,
\end{equation}
where the discretized charge densities are given by:
\begin{align}
    \tilde{\bm{\uprho}}_i(t) &= \sum_{p=1}^{N_i} w_{i,p} q_i \tilde{\mathcal{R}}_3 \left(S(\bX-\bX_{i,p}(t)\right),\label{eq:rhoi_disc} \\
    \tilde{\bm{\uprho}}_{e,gc}(t) &= \sum_{p=1}^{N_e} w_{e,p} q_e \tilde{\mathcal{R}}_3 \left(S(\bX-\bX_{e,gc,p}(t)\right),\label{eq:rhogc_disc}
\end{align}
Equation \eqref{eq:ampere_disc} has a solution only if the discrete divergence of the current density is 0 i.e.
\begin{equation} \label{eq:divJ_disc}
\tilde{\mathbb{D}} (\tilde{\arrJ}_i(t) + \tilde{\arrJ}_{e,gc}(t)) = 0,
\end{equation}
as implied by equation \eqref{eq:quasineutrality_disc}.
Just like the continuous counterpart, this condition can be obtained by taking the discrete divergence of equation \eqref{eq:ampere_disc}.
While equation \eqref{eq:QN-condition} is valid at the continuous level for the hybrid quasineutral model, its discrete counterpart equation \eqref{eq:quasineutrality_disc} is not satisfied up to machine precision as per our numerical scheme. On account of the number of particles used being finite, the numerical values of the discrete charge density are polluted by statistical sampling noise, and are therefore not exactly 0 in quasineutral simulations. In typical PIC simulations, this noise and the resultant charge density error decreases as $\mathcal{O}(1/\sqrt{N_p})$ \citep{birdsall2018plasma}. While equation \eqref{eq:quasineutrality_disc} is not enforced, we use the discretized Amp\`ere's equation \eqref{eq:ampere_disc}, which implies equation \eqref{eq:divJ_disc}, to obtain $\arrE_\perp$, as explained below in Section \ref{subsec:discreteEperpeqn}.

\subsection{Semi-discrete electric field equations}\label{sec:discreteefieldeqn}
\subsubsection{Equation for $\arrE_\perp$}\label{subsec:discreteEperpeqn}
Substituting \eqref{eq:Xdotgcdisc} into \eqref{eq:Jgc_disc} gives us the discretized version of equation \eqref{eq:Je_vgc}, given by:
\begin{equation}
    \tilde{\arr{J}}_{e,gc}(t) = \sum_{p=1}^{N_e} w_{e,p} q_e \tilde{\mathcal{R}}_2 \left(\left({V}_{\parallel,p} \frac{\bB^{\ast}(\bX_{e,gc,p})}{B_\parallel^{\ast}(\bX_{e,gc,p})}  + \frac{\bE_\perp^S(\bX_{e,gc,p})\times \bb_{eq} }{B_\parallel^{\ast}(\bX_{e,gc,p})} \right) S(\bX-\bX_{e,gc,p}(t))\right).
\end{equation}
Substituting the above equation in equation \eqref{eq:ampere_disc} gives us an asymmetric linear system that is computationally difficult to solve. An alternative approach would be to solve a discretized approximation of \eqref{eq:Eperp_eqn} given by:
\begin{multline}\label{eq:Eperp_eqn_disc}
\mu_0\sum_{p=1}^{N_e} w_{e,p} q_e \tilde{\mathcal{R}}_2 \left(
\frac{1}{B_\parallel^{\ast}(\bX_{e,gc,p})}
\left(\bE_\perp^S(\bX_{e,gc,p})\right)
S(\bX-\bX_{e,gc,p}(t))
\right)
\\
= \mathcal{B}_{eq} \times
\left[
 \mathbb{C}^\top\mathbb{H}_2\arrB - \mu_0\tilde{\arr{J}}_i - \mu_0\sum_{p=1}^{N_e} w_{e,p} q_e \tilde{\mathcal{R}}_2 \left(V_{\parallel,p} \frac{\bB^{\ast}(\bX_{e,gc,p})}{B_\parallel^{\ast}(\bX_{e,gc,p})} S(\bX-\bX_{e,gc,p}(t)) \right)
\right].
\end{multline}
Here, $\mathcal{B}_{eq} \times$ is an operator that is needed because the terms inside the large square brackets on the right hand side are not on the same discretized spaces as the term on the left hand side. This happens due to the design of the mimetic finite difference discretization used here. This operator performs a face-to-face projection accounting for the scaling between different dual face spaces and the appropriate signs according to the rules of the cross product operation. Equation \eqref{eq:Eperp_eqn_disc} is a large, sparse linear system that can be solved to obtain $\arrE$. The left-hand side of equation \eqref{eq:Eperp_eqn_disc} is a large, sparse particle `mass' matrix that can be approximated as:
\begin{equation}\label{eq:Eperp_approx}
    \sum_{p=1}^{N_e} w_{e,p} q_e\tilde{\mathcal{R}}_2 \left(\frac{1}{B_\parallel^{\ast}(\bX_{e,gc,p})}\left(\bE_\perp^S(\bX_{e,gc,p})\right)S(\bX-\bX_{e,gc,p}(t))\right) \approx \frac{\bar\rho_{e,gc}}{\bar{B_\parallel^{\ast}}} \arrE_\perp,
\end{equation}
where $\bar\rho_{e,gc}$ and $\bar{B_\parallel^{\ast}}$ can be found with appropriate interpolations. Such an approximation provides us $\arrE$ in straightforward manner, circumventing the computational cost of looping over each particle and the need to solve a linear system.

\subsubsection{Equation for $E_\parallel$}
Taking the time derivative of \eqref{eq:ampere_disc} and substituting $\fract{\arrB}{t}$ \eqref{eq:faraday_disc}, we get:
\begin{equation}\label{eq:curlcurlEdisc}
    \mathbb{C}^\top \mathbb{H}_2\mathbb{C}\arrE = - \mu_0\left(\fract{\tilde{\arr{J}}_i}{t} + \fract{\tilde{\arr{J}}_{e,gc}}{t}\right).
\end{equation}
Once we have $\arrE_\perp$ from the solution of \eqref{eq:Eperp_eqn_disc}, we only need to solve the $\parallel-$component of equation \eqref{eq:curlcurlEdisc} to obtain $E_\parallel$. The curl-curl operator above can also be written as:
\begin{equation}\label{eq:curlcurl}
(\mathbb{C}^\top\mathbb{H}_2 \mathbb{C}) = \mathbb{H}_1 \mathbb{G} \tilde{\mathbb{H}}_3 \tilde{\mathbb{D}} \mathbb{H}_1 - \mathbb{L},
\end{equation}
where $\mathbb{L}$ is the Laplacian operator. The $\parallel-$component of $\mathbb{C}^\top \mathbb{H}_2\mathbb{C}\arrE$ is given by:
\begin{equation}\label{eq:curlcurlparallel}
    (\mathbb{C}^\top \mathbb{H}_2\mathbb{C}\arrE)_\parallel = \mathbb{H}_1 \mathbb{G}_\parallel \tilde{\mathbb{H}}_3 \tilde{\mathbb{D}}_\perp \mathbb{H}_1\arrE_\perp - \mathbb{L}_\perp \arrE_\parallel.
\end{equation}
This relation helps break down $(\mathbb{C}^\top \mathbb{H}_2\mathbb{C}\arrE)_\parallel$ into the individual contributions of $\arrE_\parallel$ and $\arrE_\perp$. Taking the time derivative of \eqref{eq:Ji_disc} and using the equations \eqref{eq:Xdotidisc} and \eqref{eq:Vdotidisc}, we get:
\begin{gather}
	\fract{\tilde{\arr{J}}_i}{t} = \sum_{p=1}^{N_i} q_i w_{i,p} \tilde{\mathcal{R}}_2\left(\fract{\bV_{i,p}}{t} S(\bX-\bX_{i,p}(t)) - \bV_{i,p} \fract{\bX_{i,p}}{t}\cdot \nabla S(\bX-\bX_{i,p}(t)) \right) \notag \\
	= \sum_{p=1}^{N_i} w_{i,p} \tilde{\mathcal{R}}_2\left(\frac{q_i^2}{m_i}(\bE^S(t,\bX_{i,p}) + \bV_{i,p}\times \bB^S(t,\bX_{i,p})) S(\bX-\bX_{i,p}(t)) - q_i\bV_{i,p} \bV_{i,p}\cdot \nabla S(\bX-\bX_{i,p}(t)) \right). \label{eq:dJidt}
\end{gather}
To obtain the $\parallel-$component of $\fract{\tilde{\arr{J}}_{e,gc}}{t}$, we take the time derivative of \eqref{eq:Jgc_disc} and use the relations in equations \eqref{eq:Xdotgcdisc} and \eqref{eq:Vdotgcdisc}. We get:
\begin{gather}
	\fract{\tilde{J}_{e,gc\parallel}}{t} = \sum_{p=1}^{N_e} q_e w_{e,p} \tilde{\mathcal{R}}_2\left(\fract{{V}_{\parallel,p}}{t} S(\bX-\bX_{e,gc,p}) - {V}_{\parallel,p} \fract{\bX_{e,gc,p}}{t}\cdot \nabla S(\bX-\bX_{e,gc,p}) \right) \notag \\
	= \sum_{p=1}^{N_e} w_{e,p} \tilde{\mathcal{R}}_2\left(\frac{q_e^2}{m_e} E_\parallel^S(\bX_{e,gc,p}) S(\bX-\bX_{e,gc,p}) \right) \notag\\
	- \sum_{p=1}^{N_e} w_{e,p} \tilde{\mathcal{R}}_2\left(q_e {V}_{\parallel,p} \left({V}_{\parallel,p} \frac{\bB^{\ast}(\bX_{e,gc,p})}{B_\parallel^{\ast}(\bX_{e,gc,p})} + \frac{\bE_\perp^S(\bX_{e,gc,p})\times \bb_{eq}}{B_\parallel^{\ast}(\bX_{e,gc,p})} \right) \cdot \nabla S(\bX-\bX_{e,gc,p}) \right). \label{eq:dJgcdt}
\end{gather}
Using equations \eqref{eq:curlcurlparallel} \eqref{eq:dJidt} and \eqref{eq:dJgcdt}, the $\parallel-$component of equation \eqref{eq:curlcurlEdisc} becomes:
\begin{gather}
 - \mathbb{L}_\perp \arrE_\parallel + \mu_0\sum_{p=1}^{N_i} w_{i,p} \tilde{\mathcal{R}}_2\left(\frac{q_i^2}{m_i}\bE_\parallel^S(t,\bX_{i,p})S(\bX-\bX_{i,p})\right) 
 + \mu_0\sum_{p=1}^{N_e} w_{e,p} \tilde{\mathcal{R}}_2\left(\frac{q_e^2}{m_e} \bE_\parallel^S(\bX_{e,gc,p}) S(\bX-\bX_{e,gc,p}) \right) \notag\\
= \mu_0 \sum_{p=1}^{N_i} w_{i,p} \tilde{\mathcal{R}}_2\left(-\bV_{i,p}\times \bB^S(t,\bX_{i,p}) S(\bX-\bX_{i,p}) + q_i\bV_{i,p} \bV_{i,p}\cdot \nabla S(\bX-\bX_{i,p}) \right)_\parallel \notag \\
	+ \mu_0\sum_{p=1}^{N_e} w_{e,p} \tilde{\mathcal{R}}_2\left(q_e {V}_{\parallel,p}^2 \frac{\bB^{\ast}(\bX_{e,gc,p})}{B_\parallel^{\ast}(\bX_{e,gc,p})} \cdot \nabla S(\bX-\bX_{e,gc,p}) \right)\notag \\
	+ \mu_0\sum_{p=1}^{N_e} w_{e,p} \tilde{\mathcal{R}}_2\left(q_e {V}_{\parallel,p} \left(\frac{\bE_\perp^S(\bX_{e,gc,p})\times \bb_{eq}}{B_\parallel^{\ast}(\bX_{e,gc,p})} \right) \cdot \nabla S(\bX-\bX_{e,gc,p}) \right)\notag \\
	-\mu_0\sum_{p=1}^{N_e} w_{e,p} \tilde{\mathcal{R}}_2\left(\frac{\bB^{\ast}(\bX_{e,gc,p})}{B_\parallel^{\ast}(\bX_{e,gc,p})} \cdot \left(\frac{q_e^2}{m_e} \bE_\perp^S(\bX_{e,gc,p}) \right) S(\bX-\bX_{e,gc,p}) \right)%\notag\\
    -\mathbb{H}_1 \mathbb{G}_\parallel \tilde{\mathbb{H}}_3 \tilde{\mathbb{D}}_\perp \mathbb{H}_1\arrE_\perp. \label{eq:Epar_eqn_disc}
\end{gather}
Equation \eqref{eq:Epar_eqn_disc} is a large, sparse linear system that is solved to obtain $\arrE_\parallel$. This is the discretized version of equation \eqref{eq:Epar_eqn}. Similar to the equation for $\arrE_\perp$, the second and third terms of equation \eqref{eq:Epar_eqn_disc} are also particle `mass' matrices that can be approximated as:
\begin{align}
    \sum_{p=1}^{N_i} w_{i,p} \tilde{\mathcal{R}}_2\left(\frac{q_i^2}{m_i}\bE_\parallel^S(t,\bX_{i,p})S(\bX-\bX_{i,p})\right) &\approx \frac{q_i \bar\rho_{i}}{m_i \bar{B_\parallel^{\ast}}} \arrE_\parallel,\\
    \sum_{p=1}^{N_e} w_{e,p} \tilde{\mathcal{R}}_2\left( \frac{q_e^2}{m_e} \bE_\parallel^S(\bX_{e,gc,p}) S(\bX-\bX_{e,gc,p}) \right) &\approx \frac{q_e \bar\rho_{e,gc}}{m_e \bar{B_\parallel^{\ast}}} \arrE_\parallel,
\end{align}
reducing the computational cost of building the particle-based matrix.
%%%%%%%%%%%%%%%%%%%%%%%%%%%%%%%%%%%%%%%%%%%%%%%%%%%%%%%%%

%%%%%%%%%%%%%%%%%%%%%%%%%%%%%%%%%%%%%%%%%%%%%%%%%%%%%%%%%
\section{Cold plasma stability analysis of hybrid quasineutral model} \label{sec:stabilityanalysis}

In this section, we obtain a time-step criterion for stability of the numerical time-stepping algorithm. A linear stability analysis is performed using the simplified 1D cold plasma model. We first obtain a maximum time-step for the fully explicit RK scheme, and then for the implicit-explicit (IMEX) schemes.

\subsection{Linearization of the 1D model}

We consider linear perturbations to the equilibrium state in the cold plasma limit of the hybrid quasineutral model considered here. We assume the equilibrium magnetic field to be in the $z-$direction i.e. $\bb_{eq} = \hat{z}$. We also assume all gradients in the $x$ and $y$ directions to be zero. The linear perturbations applied are of the form:
\begin{equation}\label{eq:perturbation_cold1D}
\bE_1 = \hat{\bE} e^{i(k z -\omega t)}.
\end{equation}
Linearizing \eqref{eq:qndjdt} in this manner gives the following equations for the individual components of $\fracp{\bJ_{i,1}}{t}$:
\begin{align}
    \fracp{J_{i,x,1}}{t} &= \epsilon_0 \omega_{p,i}^2 E_{x,1} + \Omega_{c,i}J_{i,y,1}, \label{eq:jicoldx}\\
    \fracp{J_{i,y,1}}{t} &= \epsilon_0 \omega_{p,i}^2 E_{y,1} - \Omega_{c,i}J_{i,x,1}, 
    \label{eq:jicoldy} \\
    \fracp{J_{i,z,1}}{t} &= \epsilon_0 \omega_{p,i}^2 E_{z,1}. \label{eq:jicoldz}
\end{align}
Linearizing equation \eqref{eq:Je_vgc} using equation \eqref{eq:Xedot} similarly gives us:
\begin{align}
	J_{e,x,1} &=  \epsilon_0 \frac{\omega_{p,e}^2}{\Omega_{c,e}} E_{y,1}  \label{eq:jecoldx}\\
	J_{e,y,1} &=  -\epsilon_0 \frac{\omega_{p,e}^2}{\Omega_{c,e}}E_{x,1}\label{eq:jecoldy}\\
	\fracp{J_{e,z,1}}{t} &=  \epsilon_0 \omega_{p,e}^2 E_{z,1}. \label{eq:jecoldz}
\end{align}
In the above equations, the terms $\omega_{p,i}$ and $\omega_{p,e}$ are the ion and electron plasma frequency, respectively. The terms $\Omega_{c,i}$ and $\Omega_{c,e}$ are the ion and electron cyclotron frequency, respectively. These frequencies are given by:
\begin{equation}\label{eq:omegas}
\omega_{p,i} = \sqrt{\frac{Z_i^2 n_i e^2}{\epsilon_0 m_i}}, ~~~~ \omega_{p,e} = \sqrt{\frac{n_e e^2}{\epsilon_0 m_e}},
\end{equation}
and
\begin{equation}\label{eq:Omegas}
\Omega_{c,i} = \frac{e Z_i B_{eq}}{m_i}, ~~~~ \Omega_{c,e} = -\frac{e B_{eq}}{m_e},
\end{equation}
where $Z_i$ is the ion atomic number, $e$ is the electronic charge, and $B_{eq}$ is the magnetic field magnitude. Note that the electron cyclotron frequency, $\Omega_{c,e}$, is taken to be negative. The terms $n_i$ and $n_e$ are the ion and electron particle number density, respectively. Similarly, $m_i$ and $m_e$ denote the respective particle masses.

We next consider the linearization of the 1D Maxwell equations. Accordingly, equations \eqref{eq:AmpereQN} and \eqref{eq:qnfaraday} become:
\begin{gather}
    c^2\fracp{B_{y,1}}{z} = - \frac{1}{\epsilon_0} (J_{i,x,1} + J_{e,x,1}), \label{eq:JixJex} \\
    c^2\fracp{B_{x,1}}{z} = \frac{1}{\epsilon_0} (J_{i,y,1} + J_{e,y,1}), \label{eq:JiyJey} \\
    (J_{i,z,1} + J_{e,z,1}) = 0, \label{eq:JizJez} \\
    \fracp{B_{x,1}}{t} = \fracp{E_{y,1}}{z}, \label{eq:Bx1} \\
    \fracp{B_{y,1}}{t} = -\fracp{E_{x,1}}{z}, \label{eq:By1} \\
    \fracp{B_{z,1}}{t} = 0. \label{eq:Bz1}
\end{gather}
From equation \eqref{eq:Bz1}, we see that $B_{z,1}$ is constant. From equations \eqref{eq:jicoldz}, \eqref{eq:jecoldz} and \eqref{eq:JizJez}, we see that $E_{z,1} = 0$, and $J_{i,z,1}$ and $J_{e,z,1}$ are constants. From equations \eqref{eq:jecoldx} and \eqref{eq:jecoldy} we see that the electron current components $J_{e,x,1}$ and $J_{e,y,1}$ can be directly written as constant multiples of $E_{y,1}$ and $E_{x,1}$ and can therefore be eliminated from the rest of the system. The unknowns that finally remain are therefore $E_{x,1}$, $E_{y,1}$, $B_{x,1}$, $B_{y,1}$, $J_{i,x,1}$ and $J_{i,y,1}$. The six coupled equations that govern their evolution are equations \eqref{eq:jicoldx}, \eqref{eq:jicoldy}, \eqref{eq:JixJex}, \eqref{eq:JiyJey}, \eqref{eq:Bx1} and \eqref{eq:By1}. For the space discretization of these six unknowns, we use a staggered centered scheme for the $z-$derivatives, wherein the electric field components are centered at the vertices of the grid, while the current densities and magnetic field components are located at the cell centers of the grid. This is also consistent with the structure-preserving discretization of our numerical model. Consider a 1D periodic grid with $N$ cells and vertices. Note that the last vertex on the right is not used as it corresponds to the first on the left due to periodicity. We then stack the six variables $E_{x,1}$, $E_{y,1}$, $B_{x,1}$, $B_{y,1}$, $J_{i,x,1}$ and $J_{i,y,1}$ from all cells into a vector $U$ of length 6$N$.
This yields a semi-discrete system of the form:
\begin{equation}
\fract{U}{t} = A U,
\end{equation}
where $A$ is a $6N \times 6N$ skew-symmetric real matrix due to the centered space discretization. This implies that all its eigenvalues $\lambda = -i\omega$ are purely imaginary. We solve the semi-discrete system with a Runge-Kutta (RK) method. It is well known, that first and second order RK methods are not stable for imaginary eigenvalues, whereas RK3 is stable for imaginary eigenvalues provided $|\lambda |\leq\sqrt{3}$ and RK4 provided $|\lambda |\leq 2\sqrt{2}$.

Rather than computing the eigenvalues of the $6N\times 6N$ matrix $A$, we can derive the corresponding numerical dispersion relation by performing a discrete Fourier transform in space. Let us denote $\Delta z = L/N$ as the grid spacing, and $k$ as the wavenumber, where $k = \frac{2\pi j}{N\Delta z}$, $j=0,1,\ldots,N-1$. This corresponds to the discrete Fourier modes of the periodic grid. The discrete Fourier transform of the centered finite difference operator for the $z-$derivative is given by:
\begin{equation}
\fracp{}{z} \rightarrow \ii k \sinc\left(\frac{k\Delta z}{2}\right),
\end{equation}
where $\sinc(x)=\frac{\sin(x)}{x}$ and the second derivative operator by:
\begin{equation}\fracp{^2}{z^2} \rightarrow - k^2 \sinc^2\left(\frac{k\Delta z}{2}\right).
\end{equation}
We observe that this is identical to the continuous with $k$ replaced by $k\sinc\left(\frac{k\Delta z}{2}\right)$, which is the modified wavenumber due to the space discretization. The maximum
modified wavenumber is $k_\text{eff,max} = \frac{2}{\Delta z}$ at the Nyquist point $k = \frac{\pi}{\Delta z}$.

\subsection{Numerical dispersion relation}

In order to compute the numerical dispersion relation for this system, we replace $\partial/\partial t$ with $-i\omega$. Initially, we replace $\partial/\partial z$ with $ik$ for simplicity. Later, when we obtain the final equation, we can replace $k$ with $k\sinc(k \Delta z/2)$ to account for the space discretization and obtain the time-stepping criterion. We first use equations \eqref{eq:jecoldx} and \eqref{eq:jecoldy} to write $J_{e,x,1}$ and $J_{e,y,1}$ in terms of $E_{x,1}$ and $E_{y,1}$ in equations \eqref{eq:JixJex} and \eqref{eq:JiyJey}. Then using the newly obtained expressions for $E_{x,1}$ and $E_{y,1}$ from \eqref{eq:JixJex} and \eqref{eq:JiyJey} in equations \eqref{eq:Bx1} and \eqref{eq:By1}, we get:
\begin{align}
\fract{\hat{B}_{x,k}}{t} &= \frac{\Omega_{c,e}}{\omega_{p,e}^2} \left(c^2 k^2 \hat{B}_{y,k} - \frac{i k}{\epsilon_0} \hat{J}_{i,x,k} \right), \label{eq:11} \\
\fract{\hat{B}_{y,k}}{t} &= -\frac{\Omega_{c,e}}{\omega_{p,e}^2} \left(c^2 k^2 \hat{B}_{x,k} + \frac{i k}{\epsilon_0} \hat{J}_{i,y,k} \right). \label{eq:22}
\end{align}
Next, we use equations \eqref{eq:jecoldx} and \eqref{eq:jecoldy} to write $E_{x,1}$ and $E_{y,1}$ in terms of $J_{e,x,1}$ and $J_{e,y,1}$ in equations \eqref{eq:jicoldx} and \eqref{eq:jicoldy}. Observing that ${\Omega_{c,e}}/{\omega_{p,e}^2} = -{\Omega_{c,i}}/{\omega_{p,i}^2}$ for singly charged ions, we can directly substitute equations \eqref{eq:JixJex} and \eqref{eq:JiyJey} into equations \eqref{eq:jicoldx} and \eqref{eq:jicoldy} to obtain:
\begin{align}
 \frac{1}{\epsilon_0} \fract{\hat{J}_{i,x,k}}{t} &= -ik \frac{\Omega_{c,e}}{\omega_{p,e}^2} \omega_{p,i}^2 c^2 \hat{B}_{x,k}, \label{eq:33} \\
\frac{1}{\epsilon_0} \fract{\hat{J}_{i,y,k}}{t} &= -ik \frac{\Omega_{c,e}}{\omega_{p,e}^2} \omega_{p,i}^2 c^2 \hat{B}_{y,k}. \label{eq:44}
\end{align}
Now, replacing $\partial/\partial t$ with $-i\omega$, equations \eqref{eq:11}--\eqref{eq:44} become:
\begin{align}
-i\omega \hat{B}_{x,k} &= \frac{\Omega_{c,e}}{\omega_{p,e}^2} \left(c^2 k^2 \hat{B}_{y,k} - \frac{i k}{\epsilon_0} \hat{J}_{i,x,k} \right), \label{eq:1} \\
-i\omega \hat{B}_{y,k} &= -\frac{\Omega_{c,e}}{\omega_{p,e}^2} \left(c^2 k^2 \hat{B}_{x,k} + \frac{i k}{\epsilon_0} \hat{J}_{i,y,k} \right), \label{eq:2} \\
 -\frac{i\omega}{\epsilon_0} \hat{J}_{i,x,k} &= -ik \frac{\Omega_{c,e}}{\omega_{p,e}^2} \omega_{p,i}^2 c^2 \hat{B}_{x,k}, \label{eq:3} \\
-\frac{i\omega}{\epsilon_0} \hat{J}_{i,y,k} &= -ik \frac{\Omega_{c,e}}{\omega_{p,e}^2} \omega_{p,i}^2 c^2 \hat{B}_{y,k}. \label{eq:4}
\end{align}
After algebraically manipulating equations \eqref{eq:1}, \eqref{eq:2}, \eqref{eq:3} and \eqref{eq:4}, we obtain the following dispersion relation:
\begin{equation}
\omega - \frac{\Omega_{c,e}^2}{\omega_{p,e}^4\omega_{p,i}^2}  \frac{c^2k^2}{\omega} = \pm \frac{\Omega_{c,e}}{\omega_{p,e}^2} c^2 k^2.
\end{equation}
This is a quadratic equation in $\omega$, whose two solutions are given by:
\begin{align}
\omega_1 &= \frac{|\Omega_{c,e}|ck}{2\omega_{p,e}^2} \left(c k + \sqrt{c^2 k^2 + 4 \omega_{p,i}^2}\right), \\
\omega_2 &= \frac{|\Omega_{c,e}| c k}{2\omega_{p,e}^2} \left(-c k + \sqrt{c^2 k^2 + 4 \omega_{p,i}^2} \right).
\end{align}
We observe that for $k \to 0$, the solutions become:
\begin{align}
\omega_1 &\approx \frac{|\Omega_{c,e}|\omega_{p,i} }{\omega_{p,e}^2} c k = V_{A,i} k, \\
\omega_2 &\approx \frac{|\Omega_{c,e}|\omega_{p,i}}{\omega_{p,e}^2} c k = V_{A,i} k,
\end{align}
where $V_{A,i}$ is the ion Alfv\'en speed. We see that as $k \rightarrow 0$, both modes have the same leading-order frequency and the same group velocity. On the other hand, for $k \to \infty$:
\begin{align}
\omega_1 &\approx \frac{|\Omega_{c,e}|}{\omega_{p,e}^2} c^2 k^2, \\
\omega_2 &\to \frac{|\Omega_{c,e}|\omega_{p,i}^{2}}{\omega_{p,e}^2} {=\Omega_{c,i}}.
\end{align}
Now, replacing $k$ by $k\sinc\left(\frac{k\Delta z}{2}\right)$ to account for the space discretization, we observe that $\omega_1$ is unbounded when $k\to \infty$, whereas $\omega_2$ remains bounded. This implies that the eigenvalues $\lambda_1 = -\ii \omega_1$ will be the largest for the smallest $\Delta z$ or the largest value of $k$ i.e. $k_\text{eff,max}$, and will determine the stability condition of the time integrator. Using RK4, we thus need:
\begin{equation}
|\lambda_1 \Delta t| = |\omega_1|\Delta t \leq 2\sqrt{2},
\end{equation}
and using the expression of $\omega_1$ for large $k$, we obtain
the following stability condition:
\begin{equation}
\Delta t \leq \frac{\sqrt{2}}{2} \frac{\omega_{p,e}^2}{c^2|\Omega_{c,e}|} \Delta z^2.
\end{equation}
This shows that the time-step scales as $\Delta z^2$ for the hybrid drift-kinetic quasineutral model when using only explicit time-stepping schemes.

\subsection{A splitting scheme to lower the stability restriction}

The above stability condition can be very restrictive for small $\Delta z$. To alleviate this restriction, we can use a splitting scheme where we treat the stiff part of the system involving the highest frequencies implicitly, and the rest explicitly. The stiff part corresponds to the terms involving the second derivatives in $z$ in equations \eqref{eq:11}--\eqref{eq:44}.

The split system reads:
\begin{align}
\fract{\hat{B}_{x,k}}{t} &= - \frac{i k}{\epsilon_0} \frac{\Omega_{c,e}}{\omega_{p,e}^2} \hat{J}_{i,x,k}, \label{eq:11ex} \\
\fract{\hat{B}_{y,k}}{t} &= -\frac{i k}{\epsilon_0}\frac{\Omega_{c,e}}{\omega_{p,e}^2} \hat{J}_{i,y,k}, \label{eq:22ex} \\
 \frac{1}{\epsilon_0} \fract{\hat{J}_{i,x,k}}{t} &= -ik \frac{\Omega_{c,e}}{\omega_{p,e}^2} \omega_{p,i}^2 c^2 \hat{B}_{x,k}, \label{eq:33ex} \\
\frac{1}{\epsilon_0} \fract{\hat{J}_{i,y,k}}{t} &= -ik \frac{\Omega_{c,e}}{\omega_{p,e}^2} \omega_{p,i}^2 c^2 \hat{B}_{y,k}. \label{eq:44ex}
\end{align}
for the non-stiff part, and for the stiff part:
\begin{align}
\fract{\hat{B}_{x,k}}{t} &= \frac{\Omega_{c,e}}{\omega_{p,e}^2} c^2 k^2 \hat{B}_{y,k}, \label{eq:11im} \\
\fract{\hat{B}_{y,k}}{t} &= -\frac{\Omega_{c,e}}{\omega_{p,e}^2} c^2 k^2 \hat{B}_{x,k}, \label{eq:22im} \\
 \frac{1}{\epsilon_0} \fract{\hat{J}_{i,x,k}}{t} &= 0, \label{eq:33im} \\
 \frac{1}{\epsilon_0} \fract{\hat{J}_{i,y,k}}{t} &= 0. \label{eq:44im}
\end{align}
The dispersion relation of the implicit part can be derived similarly as before, yielding the following eigenvalues:
\begin{equation}
\lambda = \pm \ii \frac{\Omega_{c,e}}{\omega_{p,e}^2} c^2 k^2.
\end{equation}
These eigenvalues are purely imaginary, and thus can be treated stably with an implicit time integrator without time-step restriction. The Crank-Nicolson (CN) method can be used here to maintain the energy conservation of the semi-discrete system. The explicit part has the following dispersion relation:
\begin{equation}
\omega^2 = \frac{\Omega_{c,e}^2}{\omega_{p,e}^4} \omega_{pi}^2 c^2 k^2,
\end{equation}
yielding the eigenvalues
\begin{equation}
\lambda = \pm i \frac{\Omega_{c,e}}{\omega_{p,e}^2} {\omega_{p,i}} c k.
\end{equation}
Here again in the discrete case $k$ is replaced by $k\sinc\left(\frac{k\Delta z}{2}\right)$ and thus the time-step restriction for the explicit part, with a 4th order RK method, is now:
\begin{equation}
\Delta t \leq \frac{\sqrt{2}}{c} \frac{\omega_{p,e}^2}{|\Omega_{c,e}|\omega_{p,i}} \Delta z,
\end{equation}
which scales as $\Delta z$ instead of $\Delta z^2$ as before, thus significantly alleviating the time-step restriction for small $\Delta z$. Using this idea, we develop two implicit-explicit (IMEX) time-stepping schemes, wherein the explicit terms are advanced in time using a 4th order low-storage Runge-Kutta (LSRK) scheme, whereas the stiff, implicit terms are advanced using the CN method. The first IMEX scheme, performs a half-step CN advance of the implicit terms, followed by a full-step RK advance of the explicit terms, and finally another half-step CN advance of the implicit terms. This scheme is labelled `CN-RK-CN' for convenience. In the second IMEX scheme, labelled `RK-CN-RK', this order is flipped, applying first a half-step RK advance of explicit terms, followed by a full-step CN advance of implicit terms, and then again a half-step RK advance of explicit terms. We note that the critical time-step for stability for the `CN-RK-CN' scheme is half of that for the `RK-CN-RK' scheme, since the latter uses half of the total time-step for each of its RK updates, while the former uses the full time-step in a single RK update. For comparison purposes, we also implement a fully explicit time-stepping using only RK updates, in our framework. These schemes are described in detail in the next section.
%%%%%%%%%%%%%%%%%%%%%%%%%%%%%%%%%%%%%%%%%%%%%%%%%%%%%%%%%

%%%%%%%%%%%%%%%%%%%%%%%%%%%%%%%%%%%%%%%%%%%%%%%%%%%%%%%%%
\section{Time-stepping schemes}\label{sec:timestepping}
Initially, the perturbed magnetic field is taken to be zero, i.e. $\bB_{tot} = \bB_{eq}$. The particle positions ($\bX_{i,p}$, $\bX_{e,gc,p}$) and velocities ($\bv_{i,p}$, $\bv_{e,gc,p}$) are available. The initial $\arrE_\perp$ is calculated from equation \eqref{eq:Eperp_eqn_disc} and $\arrE_\parallel$ is calculated from equation \eqref{eq:Epar_eqn_disc}. The initial charge and current densities can then be obtained from equations \eqref{eq:rhoi_disc}, \eqref{eq:rhogc_disc}, \eqref{eq:Ji_disc} and \eqref{eq:Jgc_disc}. This completes the initialization process.

After setting up initial conditions, the temporal updates can be performed using the fully explicit LSRK scheme or the implicit-explicit schemes described below. The LSRK scheme is a basic component of each of the time-stepping methods considered here and therefore must be described first. The $s-$stage LSRK scheme used for solving ODEs of the form:
\begin{equation}
u' = F(u(t)), \quad u(0) = u_0,
\end{equation}
can be described as follows:
\begin{equation} \label{eq:LSRK_stages}
\begin{aligned}
    & \quad S_1 := u^n \\
    & \quad \text{for } i = 1:s \ \text{do} \\
    & \quad \quad S_2 := A_i S_2 + \Delta t F(S_1) \\
    & \quad \quad S_1 := S_1 + B_i S_2 \\
    & \quad \text{end} \\
    & \quad u^{n+1} = S_1.
\end{aligned}    
\end{equation}
The coefficients $A_1(=0), A_2, \dots, A_s$ and $B_1, \dots, B_s$ can be found in the works by \citet{williamson1980lsrk} and \citet{meng2025}.

\subsection{Fully explicit scheme}
For the fully explicit scheme, each full time-step update comprises the sequential stages described in equation \eqref{eq:LSRK_stages}. Within each stage of the LSRK algorithm, the following updates are performed:
\begin{enumerate}
\item $\arrE_\perp$ is calculated using equation \eqref{eq:Eperp_eqn_disc}.
\item $\arrE_\parallel$ is calculated using equation \eqref{eq:Epar_eqn_disc}.
\item Particle velocities are updated using equations \eqref{eq:Vdotgcdisc} and \eqref{eq:Vdotidisc}.
\item Particle positions are updated using equations \eqref{eq:Xdotgcdisc} and \eqref{eq:Xdotidisc}.
\item $\arrB$ is updated using equation \eqref{eq:faraday_disc}.
\end{enumerate}

\subsection{Implicit-explicit (IMEX) schemes}
The equation \eqref{eq:qnfaraday} can be written as:
\begin{equation}
\fracp{\bB}{t} + \nabla \times (\bE_\perp + \bE_\parallel) = \fracp{\bB}{t} + \nabla \times (\bE_{\perp,1} + \bE_{\perp,2} + \bE_\parallel) = {\bf 0}, \label{eq:qnfaraday_split}
\end{equation}
where $\bE_\perp$ has been written as $\bE_\perp = \bE_{\perp,1} + \bE_{\perp,2}$. Here $\bE_{\perp,1}$ and $\bE_{\perp,2}$ are the solutions of the equations:
\begin{equation}
     \mu_0 \int q_e \bE_{\perp,1} f_{e,gc} \dd v_{\parallel} \dd \mu = - \mu_0\bb_{eq} \times\bJ_i - \mu_0q_e\int \left(v_{\parallel}{\bb_{eq} \times\bB^{\ast}} - \frac{\mu}{q_e} \nabla_\perp B_{\parallel,tot} \right) f_{e,gc} \dd v_{\parallel} \dd \mu\label{eq:Eperp1_eqn}.
\end{equation}
\begin{equation}
     \mu_0 \int q_e \bE_{\perp,2} f_{e,gc} \dd v_{\parallel} \dd \mu = \bb_{eq} \times \nabla\times\bB \label{eq:Eperp2_eqn}.
\end{equation}
Equations \eqref{eq:Eperp1_eqn} and \eqref{eq:Eperp2_eqn} add up to give equation \eqref{eq:Eperp_eqn}. Note that we assume $\mu = 0$ in equation \eqref{eq:Eperp1_eqn}, and that it is shown here only for consistency with \eqref{eq:Eperp_eqn}. For the temporal update of $\arrB$ in the numerical algorithm, the $\nabla \times \bE_{\perp,2}$ term is a stiff term that imposes a stringent constraint on the time-step used. To overcome this constraint, we use an implicit-explicit (IMEX) scheme, where the equation
\begin{equation}
\fracp{\bB}{t} + \nabla \times (\bE_{\perp,1} + \bE_\parallel) = {\bf 0}, \label{eq:qnfaraday_split1}
\end{equation}
is treated explicitly using the LSRK scheme, and the equation
\begin{equation}
\fracp{\bB}{t} + \nabla \times \bE_{\perp,2} = {\bf 0}, \label{eq:qnfaraday_split2}
\end{equation}
is treated implicitly using a Crank-Nicolson scheme.
Therefore, the explicit $\bB-$update is performed as:
\begin{equation} \label{eq:faraday_disc_E1}
    \fract{\arrB}{t} + \mathbb{C}(\arrE_{\perp,1} + \arrE_\parallel) = 0,
\end{equation}
where $\arrE_{\perp,1}$ is the discretized version of $\bE_{\perp,1}$. This is obtained by solving equation \eqref{eq:Eperp_eqn_disc}, without accounting for the $\mathbb{C}^\top\mathbb{H}_2\arrB$ term on the right hand side, i.e.
\begin{multline}\label{eq:Eperp1_eqn_disc}
\mu_0\sum_{p=1}^{N_e} w_{e,p} q_e \tilde{\mathcal{R}}_2 \left(
\frac{\bE_{\perp,1}^S(\bX_{e,gc,p})}{B_\parallel^{\ast}(\bX_{e,gc,p})}
S(\bX-\bX_{e,gc,p}(t))
\right)
\\
= \mathcal{B}_{eq} \times
\left[
 - \mu_0\tilde{\arr{J}}_i - \mu_0\sum_{p=1}^{N_e} w_{e,p} q_e \tilde{\mathcal{R}}_2 \left(V_{\parallel,p} \frac{\bB^{\ast}(\bX_{e,gc,p})}{B_\parallel^{\ast}(\bX_{e,gc,p})}\right)S(\bX-\bX_{e,gc,p}(t))
\right].
\end{multline}
The implicit $\bB-$update is performed using the semi-discrete equation:
\begin{equation} \label{eq:faraday_disc_E2}
    \fract{\arrB}{t} + \mathbb{C}\arrE_{\perp,2} = 0,
\end{equation}
where $\arrE_{\perp,2}$ is the discretized version of $\bE_{\perp,2}$ given by:
\begin{equation}\label{eq:Eperp2_eqn_disc}
\mu_0\sum_{p=1}^{N_e} w_{e,p} q_e \tilde{\mathcal{R}}_2 \left(
\frac{1}{B_\parallel^{\ast}(\bX_{e,gc,p})}
\left(\bE_{\perp,2}^S(\bX_{e,gc,p})\right)
S(\bX-\bX_{e,gc,p}(t))
\right)
= \mathcal{B}_{eq} \times
\left[\mathbb{C}^\top\mathbb{H}_2\arrB
\right].
\end{equation}
Using the approximation from equation \eqref{eq:Eperp_approx}, equation \eqref{eq:faraday_disc_E2} becomes:
\begin{equation} \label{eq:faraday_disc_E2_approx}
    \fract{\arrB}{t} + \mathbb{C}\left( \frac{\bar{B_\parallel^{\ast}}}{\bar\rho_{e,gc}}
    \mathcal{B}_{eq} \times[\mathbb{C}^\top\mathbb{H}_2\arrB] \right) = 0.
\end{equation}
Discretizing this equation in time using the Crank-Nicolson method, it becomes:
\begin{equation} \label{eq:faraday_disc_E2_approx_CN}
    \frac{\arrB^{n+1} - \arrB^n}{\Delta t} + \mathbb{C}\left( \frac{\bar{B_\parallel^{\ast}}}{\bar\rho_{e,gc}}
    \mathcal{B}_{eq} \times\left[\mathbb{C}^\top\mathbb{H}_2 \left(\frac{\arrB^{n+1} + \arrB^n}{2}\right)\right] \right) = 0,
\end{equation}
where $\Delta t$ is the discrete time-step. This is a large, sparse linear system that can be solved to obtain $\arrB^{n+1}$ given $\arrB^n$. The resultant IMEX time-stepping schemes that are a combination of the LSRK and Crank-Nicolson updates are as follows:
\subsubsection{IMEX1 (CN-RK-CN)}
\begin{enumerate}
\item $\arrB$ is updated by a half time-step $\Delta t/2$ using equation \eqref{eq:faraday_disc_E2_approx_CN}.
\item One full time-step LSRK update:
\begin{enumerate}
 \item $\arrE_\perp$ is calculated using equation \eqref{eq:Eperp_eqn_disc}. $\arrE_{\perp,1}$ is calculated using equation \eqref{eq:Eperp1_eqn_disc}.
 \item $\arrE_\parallel$ is calculated using equation \eqref{eq:Epar_eqn_disc}.
 \item Particle velocities are updated using equations \eqref{eq:Vdotgcdisc} and \eqref{eq:Vdotidisc}.
 \item Particle positions are updated using equations \eqref{eq:Xdotgcdisc} and \eqref{eq:Xdotidisc}.
 \item $\arrB$ is updated using equation \eqref{eq:faraday_disc_E1}.
 \end{enumerate}
\item $\arrB$ is updated by a half time-step $\Delta t/2$ using equation \eqref{eq:faraday_disc_E2_approx_CN}.
\end{enumerate}

\subsubsection{IMEX2 (RK-CN-RK)}
\begin{enumerate}
\item One half time-step LSRK update:
\begin{enumerate}
 \item $\arrE_\perp$ is calculated using equation \eqref{eq:Eperp_eqn_disc}. $\arrE_{\perp,1}$ is calculated using equation \eqref{eq:Eperp1_eqn_disc}.
 \item $\arrE_\parallel$ is calculated using equation \eqref{eq:Epar_eqn_disc}.
 \item Particle velocities are updated using equations \eqref{eq:Vdotgcdisc} and \eqref{eq:Vdotidisc}.
\item Particle positions are updated using equations \eqref{eq:Xdotgcdisc} and \eqref{eq:Xdotidisc}.
 \item $\arrB$ is updated using equation \eqref{eq:faraday_disc_E1}.
 \end{enumerate}
\item $\arrB$ is updated by a full time-step $\Delta t$ using equation \eqref{eq:faraday_disc_E2_approx_CN}.
\item One half time-step LSRK update:
\begin{enumerate}
 \item $\arrE_\perp$ is calculated using equation \eqref{eq:Eperp_eqn_disc}. $\arrE_{\perp,1}$ is calculated using equation \eqref{eq:Eperp1_eqn_disc}.
 \item $\arrE_\parallel$ is calculated using equation \eqref{eq:Epar_eqn_disc}.
\item Particle velocities are updated using equations \eqref{eq:Vdotgcdisc} and \eqref{eq:Vdotidisc}.
\item Particle positions are updated using equations \eqref{eq:Xdotgcdisc} and \eqref{eq:Xdotidisc}.
 \item $\arrB$ is updated using equation \eqref{eq:faraday_disc_E1}.
 \end{enumerate}
\end{enumerate}
%%%%%%%%%%%%%%%%%%%%%%%%%%%%%%%%%%%%%%%%%%%%%%%%%%%%%%%%%

%%%%%%%%%%%%%%%%%%%%%%%%%%%%%%%%%%%%%%%%%%%%%%%%%%%%%%%%%
\section{Dispersion relation}\label{sec:dispersionrelation}
It is possible to perform a linear perturbation analysis of the quasineutral hybrid drift-kinetic Vlasov-Maxwell system to obtain a dispersion relation describing the various eigenmodes generated by the system. The dispersion relation can be obtained from the non-quasineutral, hybrid drift-kinetic Vlasov-Maxwell system \citep{meng2025} by simply taking the quasineutral limit i.e. $\epsilon_0 \to 0$. We now describe the dispersion relation for the system of governing equations considered here. An equilibrium state is assumed where $\bE = \bE_0$, $\bB = \bB_{eq}$, $f_e = f_{e,0}$ and $f_i = f_{i,0}$. Perturbations from equilibrium quantities are denoted by the subscript `1'. To obtain the dispersion relation for linear stability analysis, plane-wave perturbations of the form:
\begin{equation}\label{eq:perturbation}
\bE_1 = \hat{\bE} e^{i(\bk \cdot \bx-\omega t)},
\end{equation}
can be assumed, where $\bk$ is the wavenumber and $\omega$ is the angular frequency for the plane-wave perturbation. Applying such perturbations to equations \eqref{eq:Je}, \eqref{eq:Ji}, \eqref{eq:VlasovDKe}, \eqref{eq:VlasovFKi}, \eqref{eq:AmpereQN} and \eqref{eq:qnfaraday}, we get:

\begin{gather}
    \hat{\bJ}_{e,gc} = \begin{pmatrix}
		{q_e n_0 \hat E_y}/{B_0} \\ -{q_e n_0 \hat E_x}/{B_0} \\ \hat J_{e,gc\parallel}
	\end{pmatrix}, \\
    \hat{\bJ_i} = q_i \int \hat{f_i} \bv d\bv, \\
    -i \omega \hat{f_e} + i k_\parallel v_{\parallel} \hat{f_e} + \frac{q_i}{m_i}\left(\hat{E}_z    \fracp{\hat{f_e}}{v_{\parallel}} \right) =0, \\
    -i \omega \hat{f_i} + i{\bv_i}\cdot \bk \hat{f_i} + \frac{q_i}{m_i}\left[{(\bv_i \times \bB_{tot})} \cdot\fracp{\hat{f_i}}{\bv_i} + {(\hat{\bE} + \bv_i \times \hat{\bB})} \cdot\fracp{f_{i,0}}{\bv_i} \right] =0, \\
    i \bk \times \hat{\bB} = \mu_0 (\hat{\bJ}_i + \hat{\bJ}_{e,gc}), \\
    \omega\hat{\bB} = \bk \times \hat{\bE}.
\end{gather}
Eliminating $\hat{\bB}$, $\hat{f_e}$, $\hat{f_i}$, $\hat{\bJ}_{e,gc}$ and $\hat{\bJ}_i$ from these equations, we obtain the equation:
\begin{equation} \label{eq:Ehat1}
     \bk \times \bk \times \hat{\bE} + (\omega/c)^2 \underline{\underline{\boldsymbol{\epsilon}}} \cdot \hat{\bE} = 0,
\end{equation}
where $\underline{\underline{\epsilon}}$ is the dielectric tensor of the plasma, that can be written as:
\begin{equation}\label{eq:epsilon}
	\underline{\underline{\boldsymbol{\epsilon}}} = \begin{bmatrix}
   \epsilon_{xx} & \epsilon_{xy} & \epsilon_{xz} \\
   \epsilon_{yx} & \epsilon_{yy} & \epsilon_{yz} \\
   \epsilon_{zx} & \epsilon_{zy} & \epsilon_{zz}
   \end{bmatrix}.
\end{equation}
The plasma dielectric tensor is a matrix that characterizes how a magnetized plasma responds to an applied electric field. It governs the dispersion relation and encapsulates the properties of linear wave propagation in a uniform magnetized plasma. For the hybrid model considered here, the dielectric tensor can be written as a sum of the contributions of the individual species i.e. $\underline{\underline{\epsilon}} = \underline{\underline{\epsilon_e}} + \underline{\underline{\epsilon_i}}$. Denoting the magnitude of $\bk$ as $k$, we can define the refractive index $n = kc/\omega$ and the wavevector $\boldsymbol{\kappa} = \bk/k$, equation \eqref{eq:Ehat1} becomes:
\begin{equation} \label{eq:Ehat2}
     n^2 \boldsymbol{\kappa} \times \boldsymbol{\kappa} \times \hat{\bE} + \underline{\underline{\boldsymbol{\epsilon}}} \cdot \hat{\bE} = 0.
\end{equation}
Taking the determinant of this linear system gives us the hot plasma dispersion relation:
\begin{equation}\label{eq:dispreln}
\det|n^2(\kappa_i \kappa_j - \delta_{ij}) + \epsilon_{ij}| = 0.
\end{equation}
Given the equivalence of all directions perpendicular to the static background magnetic field, we write $\underline{\underline{\boldsymbol{\epsilon}}}$ in terms of $k_{\perp}$ and $k_{\parallel}$, which are the components of $\bk$ perpendicular and parallel to the magnetic field, respectively. We assume $\theta$ to be the angle between $\bk$ and the static background magnetic field. Without loss of generality, we define $k_x = k_{\perp}$, $k_y = 0$ and $k_z = k_{\parallel}$, and therefore get:
\begin{gather}
    k_x = k_{\perp} = k \sin(\theta) \Longrightarrow \kappa_x = \sin(\theta), \label{eq:k_perp} \\
    k_z = k_{\parallel} = k \cos(\theta) \Longrightarrow \kappa_z = \cos(\theta). \label{eq:k_par}
\end{gather}
The background magnetic field is therefore along the $z-$direction. The hot plasma dispersion relation, thus becomes:
\begin{equation}\label{eq:hot_dispreln}
	\det\begin{bmatrix}
   \epsilon_{xx} - n^2\cos^2(\theta) & \epsilon_{xy} & \epsilon_{xz} + n^2\cos(\theta)\sin(\theta)\\
   \epsilon_{yx} & \epsilon_{yy} - n^2 & \epsilon_{yz} \\
   \epsilon_{zx} + n^2\cos(\theta)\sin(\theta) & \epsilon_{zy} & \epsilon_{zz} - n^2\sin^2(\theta)
   \end{bmatrix} = 0.
\end{equation}
A detailed derivation of the dispersion relation for the drift-kinetic species, accounting for the nonlinear polarization and magnetization terms, can be found in the work by \citet{Zonta2021Dispersion}. A simplified model without these terms is provided in the work by \citet{meng2025}. We also define $n_{\perp} = n\sin(\theta)$ and $n_{\parallel} = n\cos(\theta)$. In the model considered here, the $\epsilon_{e,xy}$, $\epsilon_{e,yx}$ and $\epsilon_{e,zz}$ terms of the drift-kinetic electron dielectric tensor $\underline{\underline{\boldsymbol{\epsilon_e}}}$ are given by:
\begin{gather}
\epsilon_{e,xy} = -\epsilon_{e,yx} = i \frac{q_e n_e}{\epsilon_0 B_{eq}\omega} = i \frac{c^2 \omega_{c,e}}{V_{A,e}^2 \omega}, \label{eq:epsilonxy} \\
\epsilon_{e,zz} = \frac{\omega_{p,e}^2}{k_\parallel^2v_{th,e}^2} \left[1+\zeta_e Z(\zeta_e)\right], \label{eq:epsilonzz}
\end{gather}
where $\zeta_e = \frac{\omega}{\sqrt(2) k_\parallel v_{th,e}}$. The terms $\omega_{p,e}$ and $\Omega_{c,e}$ are the electron plasma frequency and electron cyclotron frequency already provided in equations \eqref{eq:omegas} and \eqref{eq:Omegas}, respectively. Also, $v_{th,e}$ is the electron thermal velocity given by:
\begin{equation}\label{eq:vth_e}
v_{th,e} = \sqrt{\frac{2T_e}{m_e}},
\end{equation}
where $T_e$ is the electron temperature. The term $V_{A,e}$ is the electron Alfv\'en speed given by:
\begin{equation}
    V_{A,e} = \frac{B_{eq}}{\sqrt{\mu_0  m_e n_e}}.
\end{equation}
The remaining $\epsilon_{e}$ terms become zero i.e. $\epsilon_{e,xx} = \epsilon_{e,xz} = \epsilon_{e,zx} = \epsilon_{e,yy} = \epsilon_{e,yz} = \epsilon_{e,zy} = 0$. The individual terms of the fully kinetic ion dielectric tensor $\underline{\underline{\boldsymbol{\epsilon_i}}}$ are given by \citet{brambilla1998kinetic}:
\begin{align}
\epsilon_{i,xx} &= - \frac{\omega_{p,i}^2}{\omega^2} \sum_{n=-\infty}^{n=+\infty} \frac{n^2}{\lambda_i} I_n(\lambda_i) e^{-\lambda_i}(-x_{0,i}Z(x_{n,i})), \label{eq:epsilonxx} \\
\epsilon_{i,xy} = -\epsilon_{i,yx} &= -i \frac{\omega_{p,i}^2}{\omega^2} \sum_{n=-\infty}^{n=+\infty} n[I'_n(\lambda_i) - I_n(\lambda_i)] e^{-\lambda_i}(-x_{0,i}Z(x_{n,i})), \label{eq:epsilonxy} \\
\epsilon_{i,xz} = \epsilon_{i,zx} &= -\half n_{\perp} n_{\parallel}  \frac{\omega_{p,i}^2}{\omega \Omega_{c,i}} \frac{v_{th,i}^2}{c^2} \sum_{n=-\infty}^{n=+\infty} \frac{n}{\lambda_i} I_n(\lambda_i) e^{-\lambda_i} (x_{0,i}^2 Z'(x_{n,i})), \label{eq:epsilonxz} \\
\epsilon_{i,yy} &= - \frac{\omega_{p,i}^2}{\omega^2} \sum_{n=-\infty}^{n=+\infty} \left[ \frac{n^2}{\lambda_i} I_n(\lambda_i) - 2\lambda_i [I'_n(\lambda_i) - I_n(\lambda_i)] \right] e^{-\lambda_i}(-x_{0,i}Z(x_{n,i})), \label{eq:epsilonyy} \\
\epsilon_{i,yz} = -\epsilon_{i,zy} &= \frac{i}{2} n_{\perp} n_{\parallel} \frac{\omega_{p,i}^2}{\omega \Omega_{c,i}} \frac{v_{th,i}^2}{c^2} \sum_{n=-\infty}^{n=+\infty} [I'_n(\lambda_i) - I_n(\lambda_i)] e^{-\lambda_i} (x_{0,i}^2 Z'(x_{n,i})), \label{eq:epsilonyz} \\
\epsilon_{i,zz} &= - \frac{\omega_{p,i}^2}{\omega^2} \sum_{n=-\infty}^{n=+\infty} I_n(\lambda_i)e^{-\lambda_i}(x_{0,i} x_{n,i} Z'(x_{n,i})). \label{eq:epsilonzz}
\end{align}
In the above expressions, $\lambda_i$ and $x_{n,i}$ are dimensionless quantities given by:
\begin{equation}\label{eq:lambda}
    \lambda_i = \frac{k_{\perp}^2 v_{th,i}^2}{2 \Omega_{c,i}^2},
\end{equation}
and
\begin{equation}\label{eq:xns}
x_{n,i} = \frac{\omega - n\Omega_{c,i}}{k_{\parallel} v_{th,i}}.
\end{equation}
The terms $\omega_{p,i}$ and $\Omega_{c,i}$ are the ion plasma frequency and ion cyclotron frequency given in equations \eqref{eq:omegas} and \eqref{eq:Omegas}, respectively, while $v_{th,i}$ is the thermal velocity given by:
\begin{equation}\label{eq:vth_i}
v_{th,i} = \sqrt{\frac{2T_i}{m_i}},
\end{equation}
where $T_i$ is the ion temperature. Also, $Z$ is the plasma dispersion function and $I_n$ denotes the modified Bessel function of order $n$.

\subsection{Cold Plasma Approximation}\label{sec:dispersionrelationcold}
To obtain a simplified dispersion relation for the case of a cold plasma, we assume the limit $T_s \rightarrow 0$. This also implies $\lambda_s \rightarrow 0$ and $|x_{n,s}| \rightarrow \infty$, causing the components $\epsilon_{xz}$, $\epsilon_{zx}$, $\epsilon_{yz}$ and $\epsilon_{zy}$ of the dielectric tensor to vanish, i.e.
\begin{equation} \label{eq:coldplasma_zeros}
\lim_{T_s \rightarrow 0} \epsilon_{xz} = \lim_{T_s \rightarrow 0} \epsilon_{zx} = \lim_{T_s \rightarrow 0} \epsilon_{yz} = \lim_{T_s \rightarrow 0} \epsilon_{zy} = 0.
\end{equation}
The remaining terms simplify as:
\begin{align}
\lim_{T_s \rightarrow 0} \epsilon_{xx} = \lim_{T_s \rightarrow 0} \epsilon_{yy} &= S = \half (R+L) = - \frac{\omega_{pi}^2}{\omega^2 - \Omega_{ci}^2}, \label{eq:coldplasma_epsilonxx} \\
\lim_{T_s \rightarrow 0} \epsilon_{xy} = -\lim_{T_s \rightarrow 0} \epsilon_{yx} &= -iD = \frac{1}{2i} (R-L) = - i \left(\frac{\omega_{pe}^2}{\omega |\Omega_{ce}|}+\frac{\Omega_{ci} \omega_{pi}^2}{\omega (\omega^2 - \Omega_{ci}^2)}\right), \label{eq:coldplasma_epsilonxy} \\
\lim_{T_s \rightarrow 0} \epsilon_{zz} &= P = - \frac{\omega_{pi}^2 + \omega_{pe}^2}{\omega^2}. \label{eq:coldplasma_epsilonzz}
\end{align}
Here the terms $R$ and $L$ are given by:
\begin{gather}
R = \frac{\omega_{pe}^2}{\omega |\Omega_{ce}|} - \frac{\omega_{pi}^2}{\omega (\omega+ \Omega_{ci})}, \label{eq:coldplasma_R} \\
L = -\frac{\omega_{pe}^2}{\omega |\Omega_{ce}|} - \frac{\omega_{pi}^2}{\omega (\omega - \Omega_{ci})}. \label{eq:coldplasma_L}
\end{gather}
Therefore, for a cold plasma, equation \eqref{eq:hot_dispreln} simplifies to:
\begin{equation}\label{eq:cold_dispreln}
	\det\begin{bmatrix}
   S - n^2\cos^2(\theta) & -iD & n^2\cos(\theta)\sin(\theta)\\
   iD & S - n^2 & 0 \\
   n^2\cos(\theta)\sin(\theta) & 0 & P - n^2\sin^2(\theta)
   \end{bmatrix} = 0.
\end{equation}
This is the cold plasma dispersion relation.

\subsection{Waves $\parallel$ to $\bB_0$}\label{sec:parallelwaves}
The dispersion relation for waves propagating in a direction parallel to the background magnetic field can be obtained by substituting $\theta = 0$ in equation \eqref{eq:cold_dispreln}. We thus get the equation:
\begin{equation}\label{eq:cold_dispreln_par}
	\det\begin{bmatrix}
   S - n^2 & -iD & 0\\
   iD & S - n^2 & 0 \\
   0 & 0 & P
   \end{bmatrix} = 0.
\end{equation}
For the quasineutral case, there is no solution to the equation $P = 0$. The remaining factor simplifies to the equations:
\begin{gather}
n^2 = S + D = R, \label{eq:coldplasma_par_R} \\
n^2 = S - D = L. \label{eq:coldplasma_par_L}
\end{gather}
For the right-handed polarized wave, the dispersion relation becomes:
\begin{equation}
    \frac{c^2k^2}{\omega^2}= R = S+D=\frac{\omega_{pe}^2}{\omega |\Omega_{ce}|} - \frac{\omega_{pi}^2}{\omega (\omega+ \Omega_{ci})}.\label{eq:upperbranch_kpar}
\end{equation}
Retaining the positive eigenfrequency, the upper right-hand branch becomes:
\begin{equation}
    \omega_R = \frac{|\Omega_{ce}|ck}{2\omega_{pe}^2} \left(c k + \sqrt{c^2 k^2 + 4 \omega_{pi}^2}\right).\label{eq:R_CAW_WHW}
\end{equation}
This is a right-handed, transverse, circularly polarized wave, called the compressional Alfv\'en wave at lower frequencies and the Whistler wave at higher frequencies (CAW-WHW). This branch has the eigenvector $(E_x, iE_x, 0)$. Similarly, for the left-handed polarized wave, the dispersion relation becomes:
\begin{equation}
    \frac{c^2k^2}{\omega^2}= L = S-D=-\frac{\omega_{pe}^2}{\omega |\Omega_{ce}|} - \frac{\omega_{pi}^2}{\omega (\omega - \Omega_{ci})}.\label{eq:lowerbranch_kpar}
\end{equation}
Retaining the positive eigenfrequency, the lower left-hand branch becomes:
\begin{equation}
    \omega_L = \frac{|\Omega_{ce}|ck}{2\omega_{pe}^2} \left(-c k + \sqrt{c^2 k^2 + 4 \omega_{pi}^2}\right).\label{eq:L_ICW}
\end{equation}
This is a left-handed, transverse, circularly polarized wave and the corresponding eigenvector is $(E_x, -iE_x, 0)$. This is called the ion cyclotron wave (ICW).

The above waves have been obtained from the cold plasma dispersion relation. However, there are other modes that do not exist in the cold plasma limit. Describing these waves requires the more generic hot plasma dispersion relation given in equation \eqref{eq:hot_dispreln}. Substituting $k_{\perp} = 0$ in the plasma dielectric tensor, the terms $\epsilon_{xz}$, $\epsilon_{yz}$, $\epsilon_{zx}$ and $\epsilon_{zy}$ simplify to 0, while the other terms are given by:
\begin{align}
\epsilon_{xx} = \epsilon_{yy} &= \half \frac{\omega_{p,i}^2}{\omega k_{\parallel} v_{th,i}} \left[Z\left(\frac{\omega - \Omega_{c,i}}{k_{\parallel} v_{th,i}}\right) + Z\left(\frac{\omega + \Omega_{c,i}}{k_{\parallel} v_{th,i}}\right) \right], \label{eq:epsilonxx_par} \\
\epsilon_{xy} = -\epsilon_{yx} &= \frac{i}{2} \frac{\omega_{p,i}^2}{\omega k_{\parallel} v_{th,i}} \left[Z\left(\frac{\omega - \Omega_{c,i}}{k_{\parallel} v_{th,i}}\right) - Z\left(\frac{\omega + \Omega_{c,i}}{k_{\parallel} v_{th,i}}\right) \right] + i \frac{c^2 \Omega_{c,e}}{V_{A,e}^2 \omega}, \label{eq:epsilonxy_par} \\
\epsilon_{zz} &= - \frac{\omega_{p,i}^2}{(k_{\parallel} v_{th,i})^2} Z'\left(\frac{\omega}{k_{\parallel} v_{th,i}}\right) + \frac{\omega_{pe}^2}{k_\parallel^2v_{th,e}^2} \left[1+\zeta_e Z(\zeta_e)\right]. \label{eq:epsilonzz_par}
\end{align}
The dispersion relation becomes:
\begin{equation}\label{eq:hot_dispreln_par}
	((\epsilon_{xx}-n^2)(\epsilon_{yy}-n^2) - \epsilon_{xy} \epsilon_{yx})\epsilon_{zz} = 0.
\end{equation}
Besides the warm plasma generalizations of the waves described above, the other solutions of this equation are the heavily damped, higher-order modes of Alfv\'en-cyclotron waves that have been studied by various researchers such as \citet{araneda2012interactions}, \citet{astudillo1996high} and \citet{Matsuda_1986}. These modes, with the $E_x$ and $E_y$ components, manifest as straight lines emanating from ($k$, $\omega$) = (0, $\Omega_{c,i}$), forming a cone-shaped structure on the $k-\omega$ plot. The mode obtained from $\epsilon_{zz} = 0$, with eigenvector $(0,0,E_z)$ also forms a similar cone-shaped structure emanating from ($k$, $\omega$) = (0, 0). These higher-order modes are a thermal effect wherein the cone-angle decreases with temperature, reducing to zero in the cold plasma limit.

\subsection{Waves $\perp$ to $\bB_0$}\label{sec:perpendicularwaves}
The dispersion relation for waves propagating in a direction perpendicular to the background magnetic field can be obtained by substituting $\theta = \pi/2$ in equation \eqref{eq:cold_dispreln}. We thus get the equation:
\begin{equation}\label{eq:cold_dispreln_perp}
	\det\begin{bmatrix}
   S & -iD & 0\\
   iD & S - n^2 & 0 \\
   0 & 0 & P - n^2
   \end{bmatrix} = 0.
\end{equation}
Again, for the quasineutral case, there is no solution for $P-n^2 = 0$. The remaining factor simplifies to:
\begin{equation}\label{eq:cold_plasma_perp}
	n^2 = \frac{RL}{S}.
\end{equation}
Substituting the expressions for $R$, $L$ and $S$, and using the relation ${\Omega_{c,e}}/{\omega_{p,e}^2} = -{\Omega_{c,i}}/{\omega_{p,i}^2}$ for singly charged ions, the above equation becomes:
\begin{equation}\label{eq:cold_plasma_perp_soln}
	n^2 = \pm\frac{\omega_{p,e}^4}{\omega_{p,i}^2 |\Omega_{c,e}|^2}.
\end{equation}
Using the definitions of the plasma and cyclotron frequencies, as provided in equations \eqref{eq:omegas} and \eqref{eq:Omegas}, and discarding the negative root, the eigenfrequency obtained is $\omega_R = V_{A,i} k$, where $V_{A,i}$ is the ion Alfv\'en speed given by:
\begin{equation} \label{eq:ion_alfven}
    V_{A,i} = \frac{B_{eq}}{\sqrt{\mu_0  m_i n_i}}.
\end{equation}
This corresponds to the Alfv\'en wave dispersion relation. The eigenvector for this wave is $(E_x, -iE_x, 0)$.

We now use the hot plasma dispersion relation to look at waves that do not exist in the cold plasma limit. Substituting $k_{\parallel} = 0$ in the plasma dielectric tensor, the terms $\epsilon_{xz}$, $\epsilon_{yz}$, $\epsilon_{zx}$ and $\epsilon_{zy}$ simplify to 0, while the other terms, including the contributions of both species, are given by:
\begin{align}
\epsilon_{xx} &= -\frac{\omega_{p,i}^2}{\omega} \frac{e^{-\lambda_i}}{\lambda_i} \sum_{n=-\infty}^{n=+\infty} \frac{n^2 I_n(\lambda_i)}{\omega - n \Omega_{c,i}}, \label{eq:epsilonxx_perp} \\
\epsilon_{xy} = -\epsilon_{yx} &= -i \frac{\omega_{p,i}^2}{\omega} e^{-\lambda_i}  \sum_{n=-\infty}^{n=+\infty} \frac{n[I'_n(\lambda_i) - I_n(\lambda_i)]}{\omega - n \Omega_{c,i}} + i \frac{c^2 \omega_{ce}}{V_{A,e}^2 \omega}, \label{eq:epsilonxy_perp} \\
\epsilon_{yy} &= -\frac{\omega_{p,i}^2}{\omega} \frac{e^{-\lambda_i}}{\lambda_i} \sum_{n=-\infty}^{n=+\infty} \frac{n^2 I_n(\lambda_i) + 2\lambda_i^2 I_n(\lambda_i) - 2\lambda_i^2 I'_n(\lambda_i)}{\omega - n \Omega_{c,i}}, \label{eq:epsilonyy_perp} \\
\epsilon_{zz} &= - \frac{\omega_{p,i}^2}{\omega} e^{-\lambda_i}  \sum_{n=-\infty}^{n=+\infty} \frac{I_n(\lambda_i)}{\omega - n \Omega_{c,i}} - \frac{\omega_{pe}^2}{\omega^2}. \label{eq:epsilonzz_perp}
\end{align}
The dispersion relation becomes:
\begin{equation}\label{eq:hot_dispreln_perp}
	(\epsilon_{xx} (\epsilon_{yy}-n^2) - \epsilon_{xy} \epsilon_{yx})(\epsilon_{zz}-n^2) = 0.
\end{equation}
We first consider the equation $(\epsilon_{zz}-n^2) = 0$, with eigenvector $(0, 0, E_z)$. From equation \eqref{eq:epsilonzz_perp}, we see that the refractive index $n = n_{\perp}$ becomes infinite at all harmonics of the ion cyclotron frequencies i.e. $\omega = m  \Omega_{c,i}$, for a positive integer $m$. Besides these cyclotron harmonic resonances, there are no wave solutions to this equation. The other factor, i.e. $(\epsilon_{xx} (\epsilon_{yy}-n^2) - \epsilon_{xy} \epsilon_{yx}) = 0$, also gives rise to these resonances on account of the $(\omega - n\Omega_{c,i})$ terms in the denominators seen on the RHS of equations \eqref{eq:epsilonxx_perp}--\eqref{eq:epsilonyy_perp}. The other solutions to this equation are the Bernstein waves \citep{bernstein} with the associated eigenvector $(E_x, E_y, 0)$. We note that the $x-$direction is the direction of wave propagation, and therefore the $E_x$ Bernstein waves are longitudinal in nature, while the $E_y$ Bernstein waves are transverse.

%%%%%%%%%%%%%%%%%%%%%%%%%%%%%%%%%%%%%%%%%%%%%%%%%%%%%%%%%

%%%%%%%%%%%%%%%%%%%%%%%%%%%%%%%%%%%%%%%%%%%%%%%%%%%%%%%%%
\section{Numerical tests}\label{sec:tests}

Tests are now conducted to validate the numerical algorithm. We consider a quasineutral plasma that is initially uniform with each species being described by a Gaussian distribution. The particles are generated using quasi-random Sobol sampling. A reduced mass ratio of $m_i/m_e = 4$ is used for all simulations considered here. Unless otherwise specified, a total of 500 particles per cell are used everywhere. The initial magnetic field is constant and uniform, and is aligned with one of the cardinal directions of the Cartesian grid. These simulations are allowed to run for a long time and the Fast-Fourier Transform (FFT) of the resultant electric field time-series data is obtained. The numerical waves obtained from these simulations are manifested on the $k-\omega$ plot of the FFT spectrum. These are plotted alongside their analytical counterparts obtained from the dispersion relation described earlier in Section \ref{sec:dispersionrelation}. We use a  quasi-1D computational domain with domain size $[0,10]\times[0,1]\times[0,1]$ and a computational grid comprising $256\times8\times8$ cells. The higher 256-cell resolution is taken in the direction along which the wave spectrum is studied. All boundaries are assumed to be periodic. A normalized background magnetic field of magnitude $B_{eq} = 1$ is used in all simulations.

\subsection{Waves propagating $\parallel$ to $\bB_0$}\label{sec:sim_parallelwaves}

\begin{figure}
    \centering
    % First row
    \begin{subfigure}[b]{0.49\textwidth}
        \centering
        \includegraphics[width=\linewidth, trim=15pt 0pt 20pt 0pt, clip]{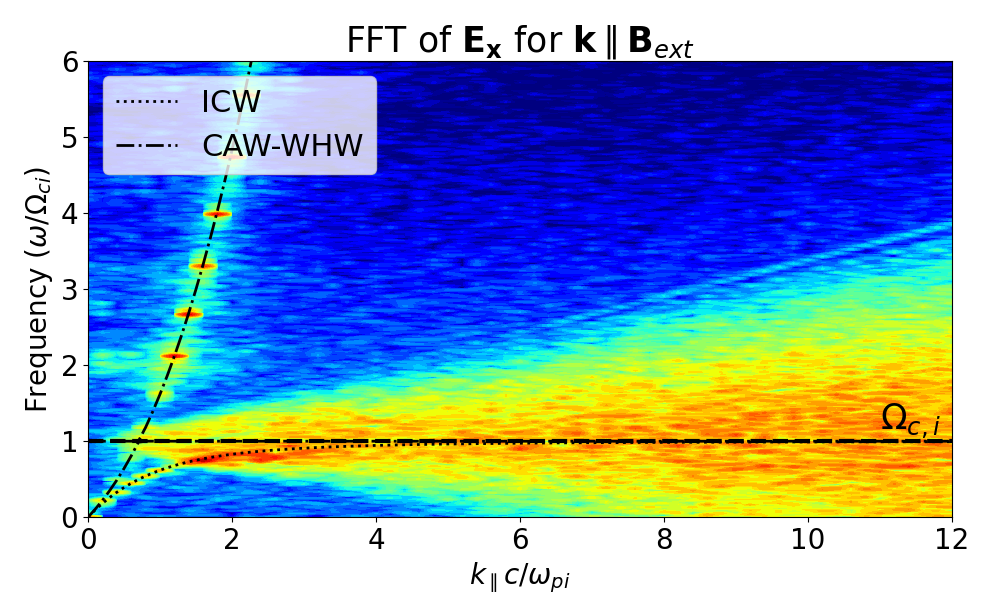}
        \caption{}
        \label{fig:kpar_rkcnrk_Ex}
    \end{subfigure}
    \hfill
    \begin{subfigure}[b]{0.49\textwidth}
        \centering
        \includegraphics[width=\linewidth, trim=15pt 0pt 20pt 0pt, clip]{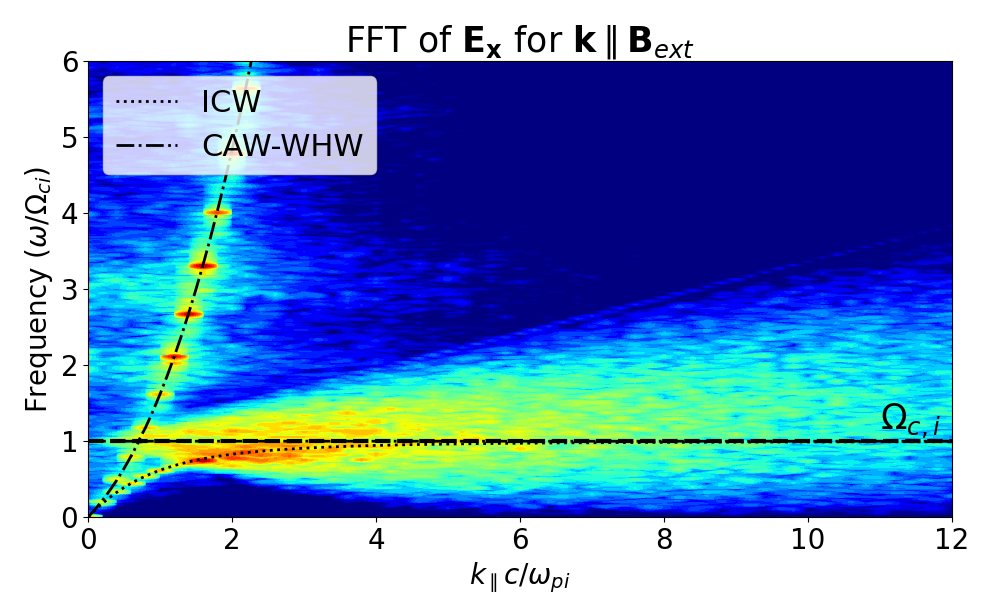}
        \caption{}
        \label{fig:kpar_rk_Ex}
    \end{subfigure}
    \hfill
    \begin{subfigure}[b]{0.49\textwidth}
        \centering
        \includegraphics[width=\linewidth, trim=15pt 0pt 20pt 0pt, clip]{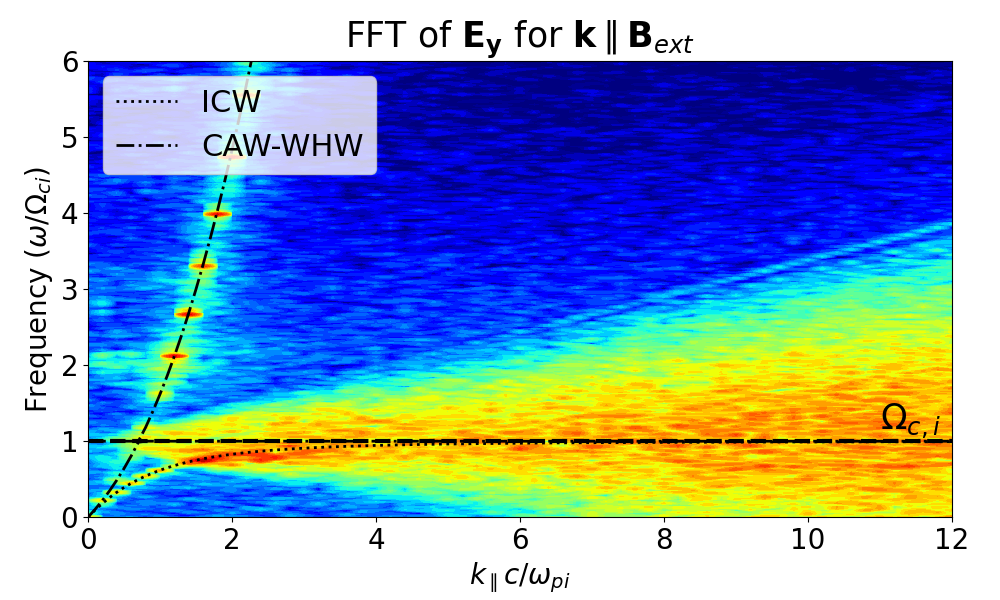}
        \caption{}
        \label{fig:kpar_rkcnrk_Ey}
    \end{subfigure}
    \hfill
    \begin{subfigure}[b]{0.49\textwidth}
        \centering
        \includegraphics[width=\linewidth, trim=15pt 0pt 20pt 0pt, clip]{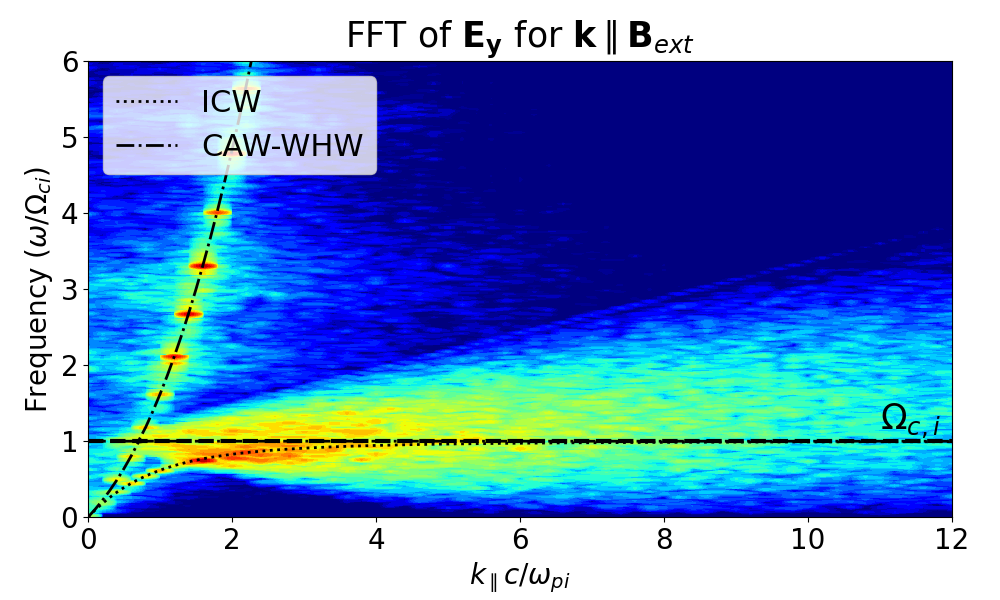}
        \caption{}
        \label{fig:kpar_rk_Ey}
    \end{subfigure}
    \hfill
    \begin{subfigure}[b]{0.49\textwidth}
        \centering
        \includegraphics[width=\linewidth, trim=15pt 0pt 20pt 0pt, clip]{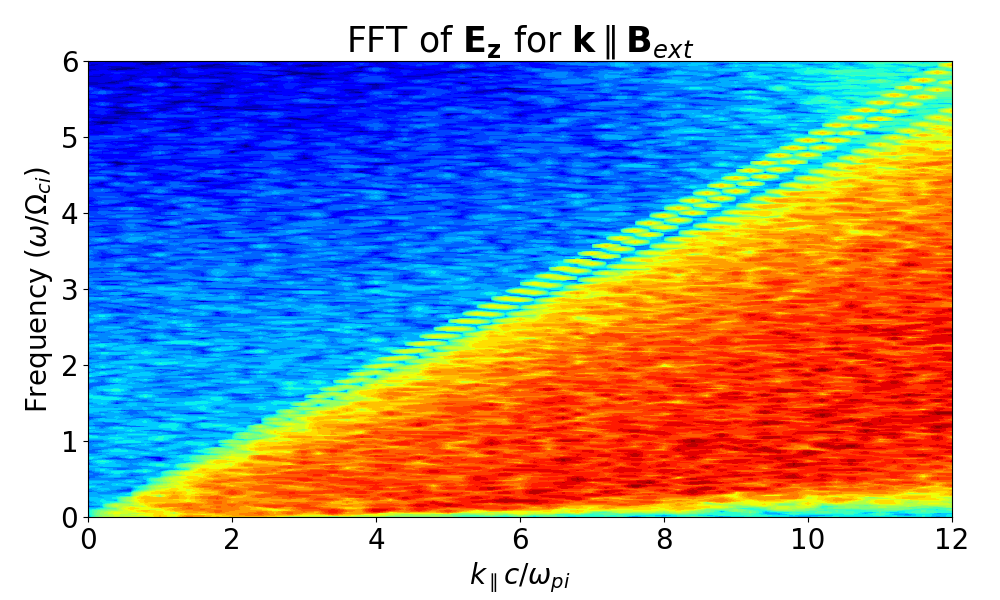}
        \caption{}
        \label{fig:kpar_rkcnrk_Ez}
    \end{subfigure}
    \hfill
    \begin{subfigure}[b]{0.49\textwidth}
        \centering
        \includegraphics[width=\linewidth, trim=15pt 0pt 20pt 0pt, clip]{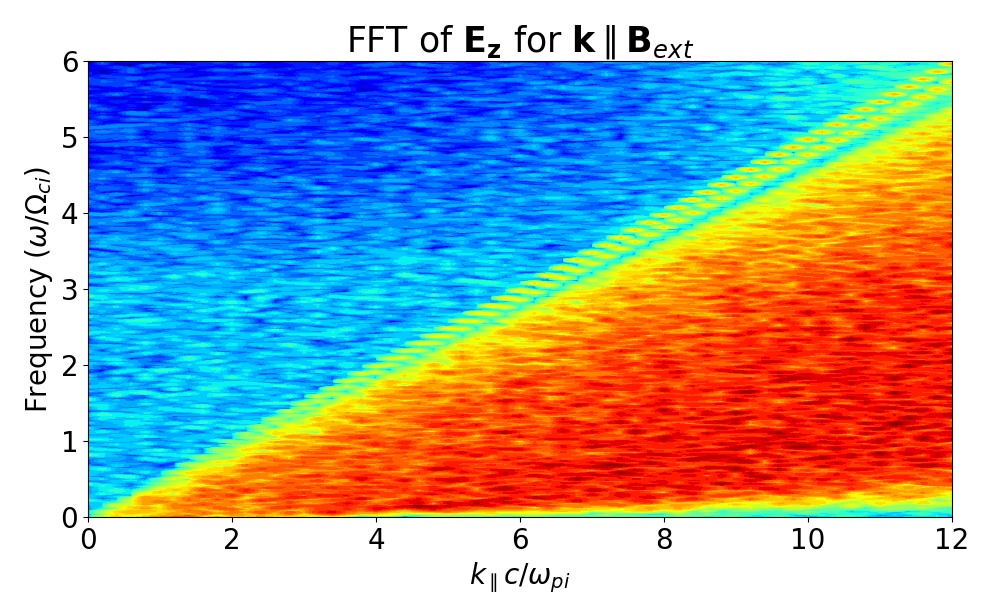}
        \caption{}
        \label{fig:kpar_rk_Ez}
    \end{subfigure}
    \caption{Wave spectra of (a) $E_x$, (c) $E_y$ and (e) $E_z(E_\parallel)$ for the `RK-CN-RK' scheme, and of (b) $E_x$, (d) $E_y$ and (f) $E_z(E_\parallel)$ for the fully explicit scheme, for $v_{th,e} = 0.05c$ and $v_{th,i} = 0.025c$, for $\bk \parallel \bB_{eq}$. $E_x$ and $E_y$ waves are transverse and $E_z(E_\parallel)$ waves are longitudinal. In (a), (b), (c) and (d), the dotted line shows the ion cyclotron wave (ICW) while the dash-dot line shows the compressional Alfv\'en-whistler wave (CAW-WHW) branch.}
    \label{fig:kpar}
\end{figure}

\begin{figure}
    \centering
    % First row
    \begin{subfigure}{0.5\textwidth}
        \centering
        \includegraphics[width=\linewidth, trim=3pt 0pt 20pt 0pt, clip]{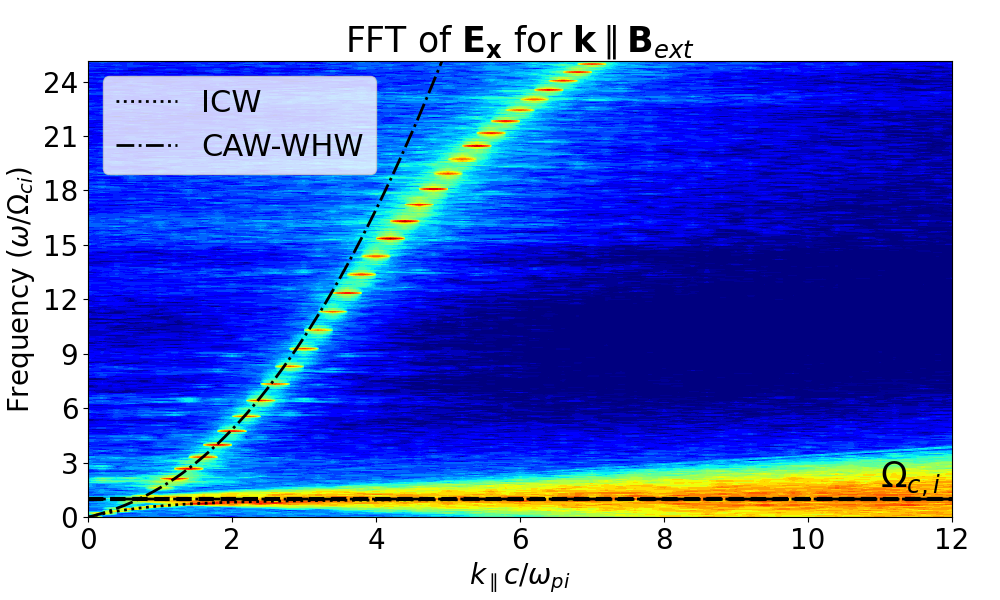}
        \caption{}
        \label{fig:kpar_rkcnrk_Ex_mag}
    \end{subfigure}
    \hfill
    % Second row
    \begin{subfigure}{0.5\textwidth}
        \centering
        \includegraphics[width=\linewidth, trim=3pt 0pt 20pt 0pt, clip]{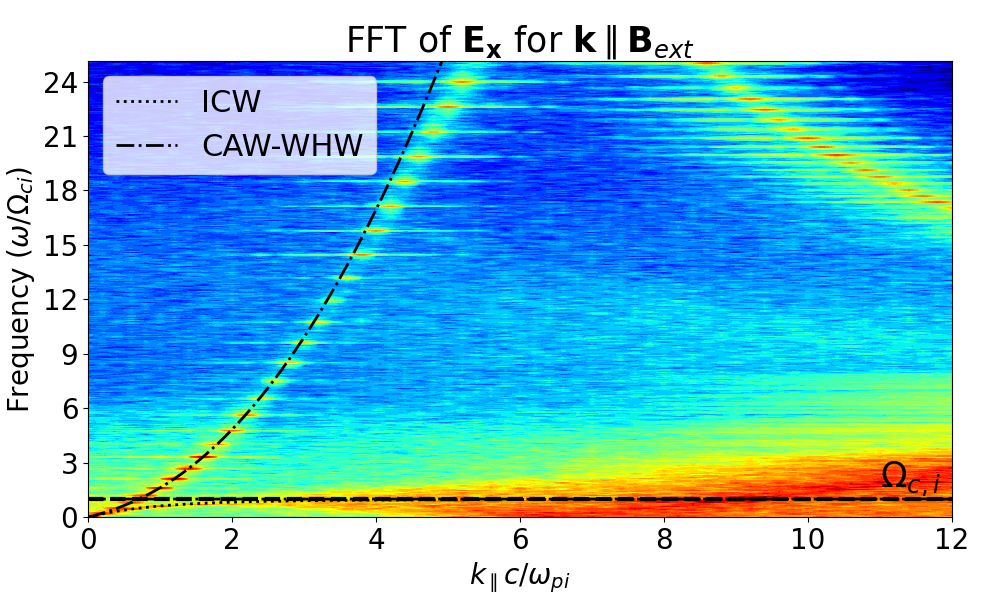}
        \caption{}
        \label{fig:kpar_cnrkcn_Ex_mag}
    \end{subfigure}
    \hfill
    \begin{subfigure}{0.5\textwidth}
        \centering
        \includegraphics[width=\linewidth, trim=3pt 0pt 20pt 0pt, clip]{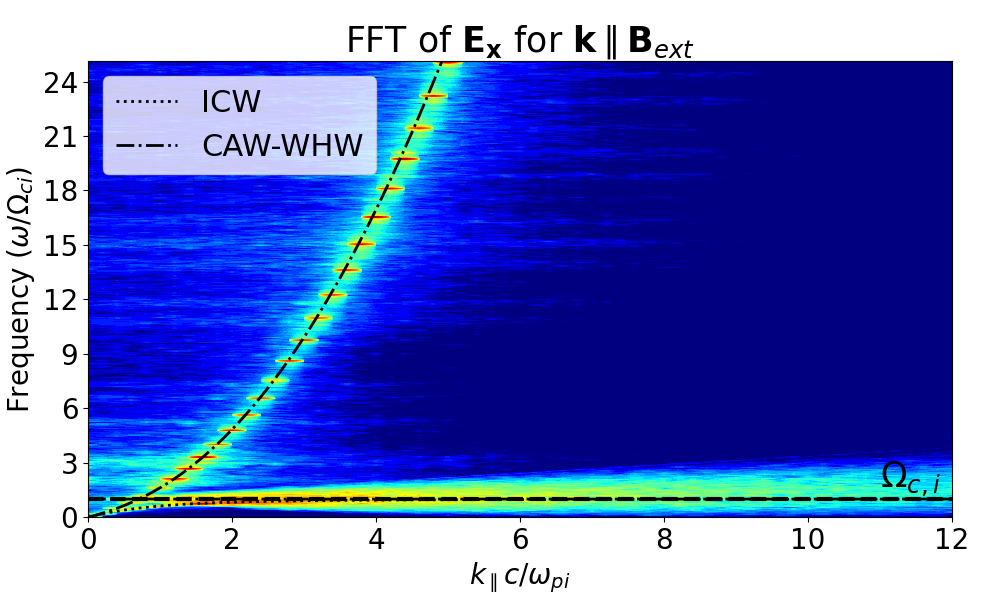}
        \caption{}
        \label{fig:kpar_rk_Ex_mag}
    \end{subfigure}
    \caption{Zoomed out plots of the $E_x$ wave spectra for $v_{th,e} = 0.05c$ and $v_{th,i} = 0.025c$, for $\bk \parallel \bB_{eq}$, obtained using the (a) `RK-CN-RK', (b) `CN-RK-CN' and (c) fully explicit schemes. The dotted line shows the ion cyclotron wave (ICW) while the dash-dot line shows the compressional Alfv\'en-whistler wave (CAW-WHW) branch.}
    \label{fig:kpar_Ex_mag}
\end{figure}

We first consider propagation of waves in the direction parallel to that of the equilibrium background magnetic field. The thermal velocity of the electrons is taken to be $v_{th,e} = 0.05c$. In order to satisfy the requirement of thermal equilibrium, the ion thermal velocity is taken to be $v_{th,i} = 0.025c$. This is calculated based on equating the temperatures of both species, and accounting for the reduced particle mass ratio. With the purpose of studying the spectrum of waves parallel to the direction of the background magnetic field, the 256-cell resolution is along this direction. The left hand side subfigures in Figure \ref{fig:kpar} show the wave spectra obtained for this case using the `RK-CN-RK' scheme, while the right hand side subfigures show those obtained using the fully explicit scheme. The results obtained using the `CN-RK-CN' are not shown for brevity. The critical time-step for numerical stability is $\Delta t_{crit} \approx 0.85$ for the `RK-CN-RK' scheme and $\Delta t_{crit} \approx 0.026$ for the fully explicit scheme. The time-steps used in these simulations are $\Delta t = 0.75$ for the `RK-CN-RK' scheme and $\Delta t = 0.02$ for the fully explicit scheme. This results in computational savings of $\sim95 \%$ when using the IMEX schemes as opposed to the fully explicit. The ion cyclotron wave (ICW) is clearly observed here in the $E_x$ and $E_y$ spectra, with the analytical result from equation \eqref{eq:L_ICW} and the ion cyclotron frequency $\Omega_{c,i}$ superimposed on it. The compressional Alfv\'en-Whistler wave (CAW-WHW) branch is also observed in the $E_x$ and $E_y$ spectra, with the analytical result from equation \eqref{eq:R_CAW_WHW} superimposed on it. The conical structure of the damped higher-order Alfv\'en-cyclotron modes described in Section \ref{sec:parallelwaves} is also clearly visible in the wave spectra of all three field components. To observe the behaviour of the time-stepping schemes at higher frequencies, zoomed out views of $E_x$ spectra obtained using all the three schemes are shown in Figure \ref{fig:kpar_Ex_mag}. On account of the small time-step used, the fully explicit method shows a strong fit with the analytical CAW-WHW branch at high frequencies as seen in Figure \ref{fig:kpar_rk_Ex_mag}. On the other hand, the IMEX schemes show a poor match with analytical results at higher frequencies. Although both IMEX schemes are formally second-order accurate and time-symmetric Strang splittings, they are not equivalent in their spectral properties. Particularly, the wave-frequency spectrum obtained from the `CN-RK-CN' simulation as shown in Figure \ref{fig:kpar_cnrkcn_Ex_mag} displays a nonphysical folding (spurious dispersion relation). This distortion arises when the non-stiff terms are evolved explicitly over a full time-step and the stiff terms are evolved implicitly over two half time-steps, introducing substantial numerical phase errors at high wavenumbers. In dispersive systems where the stiff linear operator dominates the phase evolution, advancing this operator implicitly over the full time-step is essential for preserving the correct dispersion relation.

\subsection{Waves propagating $\perp$ to $\bB_0$}\label{sec:sim_perpendicularwaves}
\begin{figure}
    \centering
    % First row
    \begin{subfigure}[b]{0.49\textwidth}
        \centering
        \includegraphics[width=\linewidth, trim=15pt 0pt 20pt 0pt, clip]{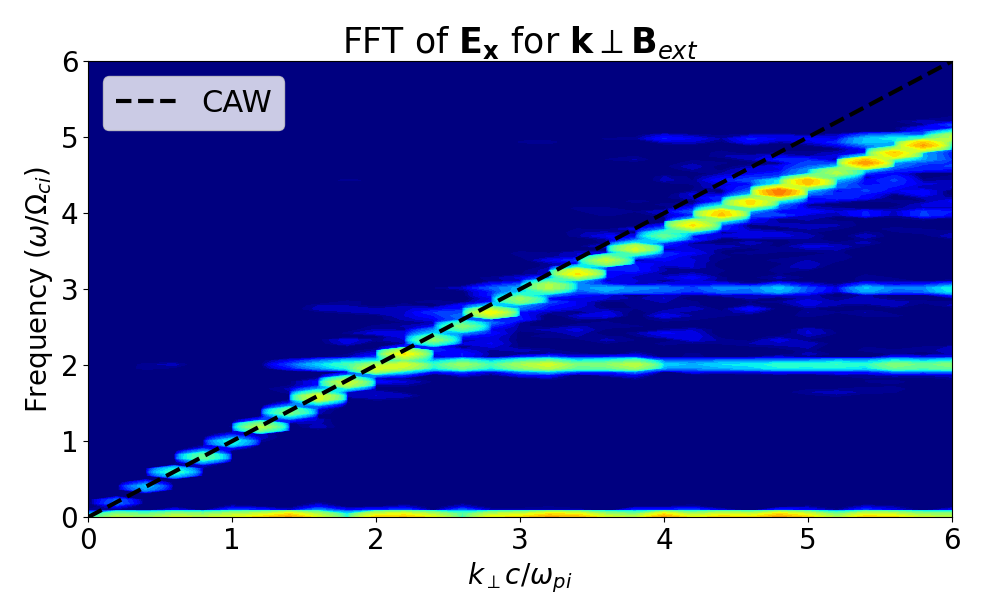}
        \caption{}
        \label{fig:kperp_cold_Ex}
    \end{subfigure}
    \hfill
    \begin{subfigure}[b]{0.49\textwidth}
        \centering
        \includegraphics[width=\linewidth, trim=15pt 0pt 20pt 0pt, clip]{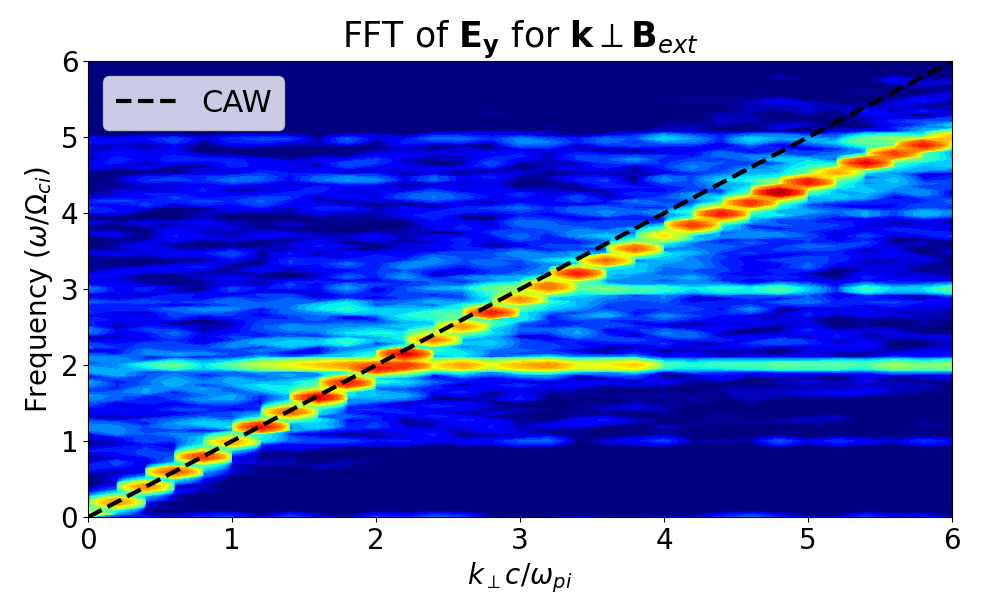}
        \caption{}
        \label{fig:kperp_cold_Ey}
    \end{subfigure}
\caption{Wave spectra of (a) $E_x$, (c) $E_y$ for $v_{th,e} = 0.05c$ and $v_{th,i} = 0.025c$, for an electron-ion plasma, for $\bk \perp \bB_{eq}$. $E_x$ waves are longitudinal and $E_y$ waves are transverse. The dashed line represents the ion compressional Alfv\'en wave, with the Alfv\'en speed given in equation \eqref{eq:ion_alfven}.}
\label{fig:kperp_cold}
\end{figure}

We now study waves propagating in a direction perpendicular to the background magnetic field. The 256-cell resolution is now along a direction perpendicular to the background magnetic field, namely the $x-$direction, as this is the direction of wave propagation studied. We first consider a cold plasma with electron thermal velocity $v_{th,e} = 0.01 c$ and ion thermal velocity $v_{th,i} = 0.005 c$. The wave spectra obtained for $E_x$ and $E_y$ are shown in Figure \ref{fig:kperp_cold}. This result shown was obtained using the `CN-RK-CN' scheme, with a time-step of $\Delta t = 0.1$. While the critical time-step as dictated by the stability criterion is $\Delta t_{crit} \approx 0.42$, the lower time-step was used to obtain a better match between the numerical and analytical dispersion relations at lower frequencies. The compressional Alfv\'en wave is clearly observed, and shows excellent agreement at lower $k$ and $\omega$ values with the theoretical prediction obtained from the cold-plasma dispersion relation. At higher $k$ and $\omega$, the numerical result shows significant deviation from the theoretical prediction, on account of the larger time-step used, as permitted by the IMEX schemes. These higher frequencies can be resolved better by using smaller time-steps, at the expense of computational efficiency, or with the use of higher-order Hodge operators. We also observe ion cyclotron resonances in these spectra, although the temperature is very low. This is because our numerical model is not a cold-plasma model, and can detect these resonances as long at the temperature is not exactly zero.

\begin{figure}
    \centering
    % First row
    \begin{subfigure}[b]{0.49\textwidth}
        \centering
        \includegraphics[width=\linewidth, trim=15pt 0pt 20pt 0pt, clip]{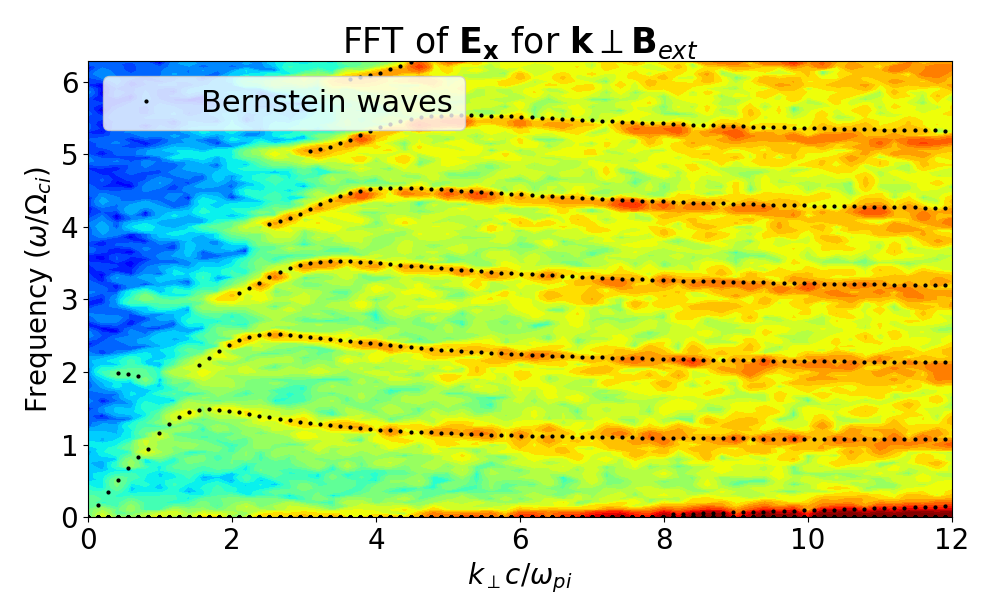}
        \caption{}
        \label{fig:kperp_hot_rkcnrk_Ex}
    \end{subfigure}
    \hfill
    \begin{subfigure}[b]{0.49\textwidth}
        \centering
        \includegraphics[width=\linewidth, trim=15pt 0pt 20pt 0pt, clip]{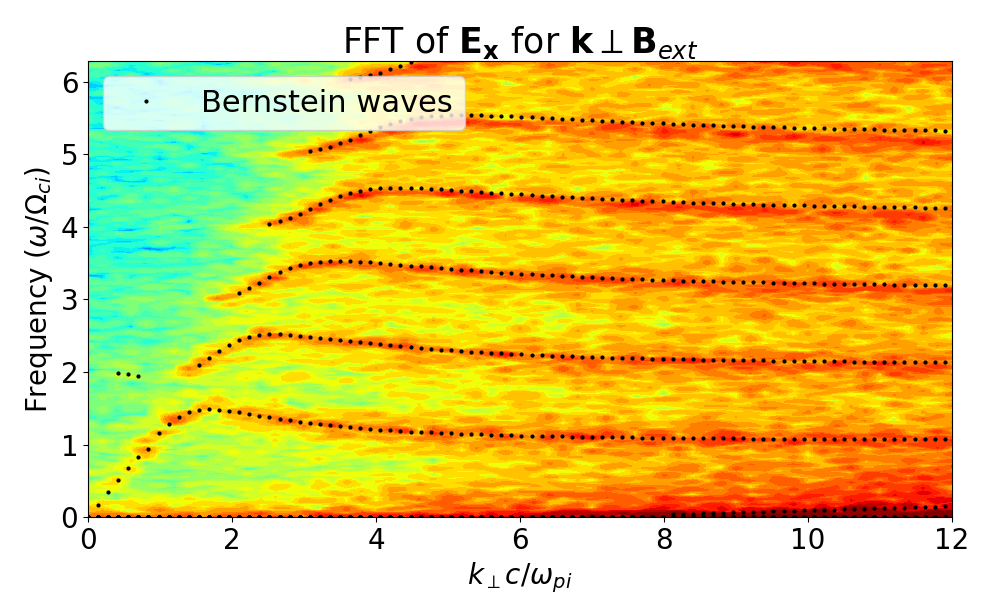}
        \caption{}
        \label{fig:kperp_hot_rkonly_Ex}
    \end{subfigure}
    \hfill
    \begin{subfigure}[b]{0.49\textwidth}
        \centering
        \includegraphics[width=\linewidth, trim=15pt 0pt 20pt 0pt, clip]{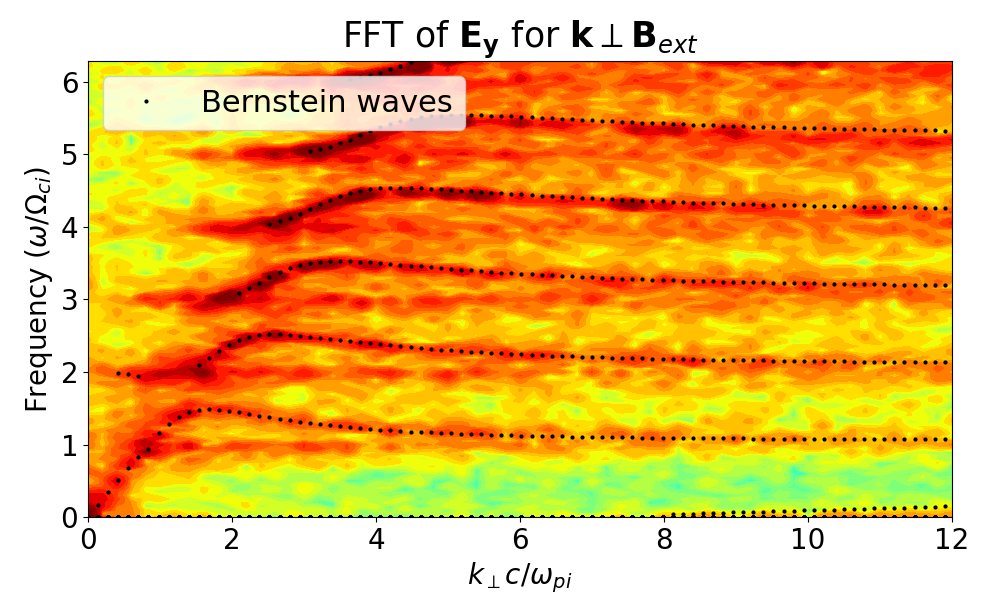}
        \caption{}
        \label{fig:kperp_hot_rkcnrk_Ey}
    \end{subfigure}
    \hfill
    \begin{subfigure}[b]{0.49\textwidth}
        \centering
        \includegraphics[width=\linewidth, trim=15pt 0pt 20pt 0pt, clip]{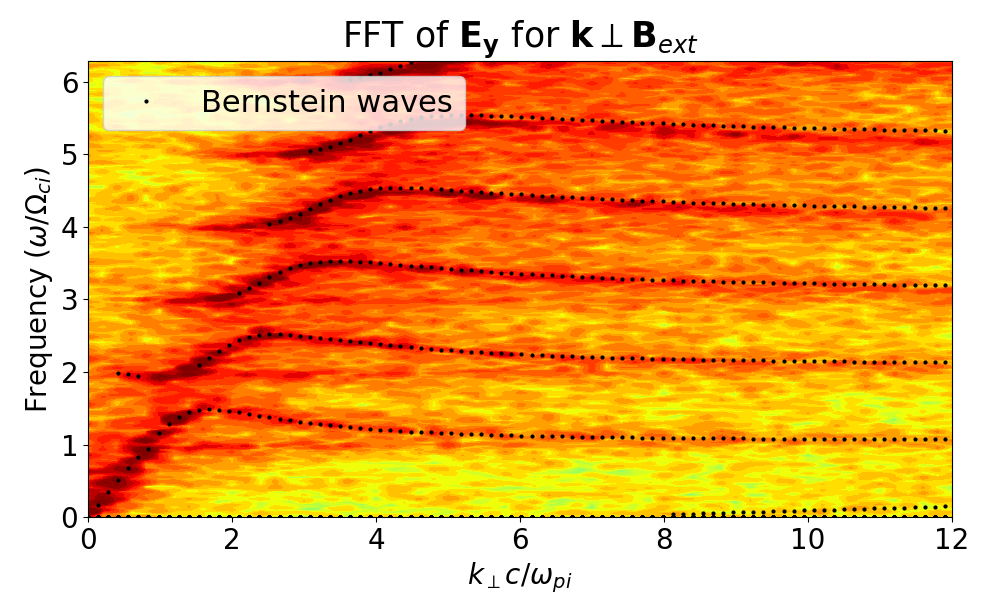}
        \caption{}
        \label{fig:kperp_hot_rkonly_Ey}
    \end{subfigure}
    \hfill
    \begin{subfigure}[b]{0.49\textwidth}
        \centering
        \includegraphics[width=\linewidth, trim=15pt 0pt 20pt 0pt, clip]{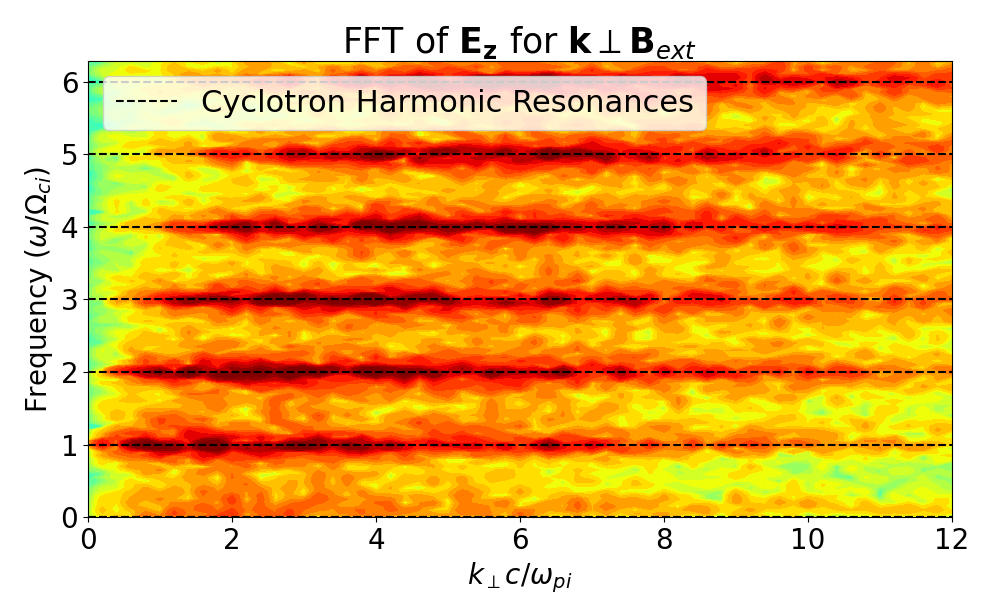}
        \caption{}
        \label{fig:kperp_hot_rkcnrk_Ez}
    \end{subfigure}
    \hfill
    \begin{subfigure}[b]{0.49\textwidth}
        \centering
        \includegraphics[width=\linewidth, trim=15pt 0pt 20pt 0pt, clip]{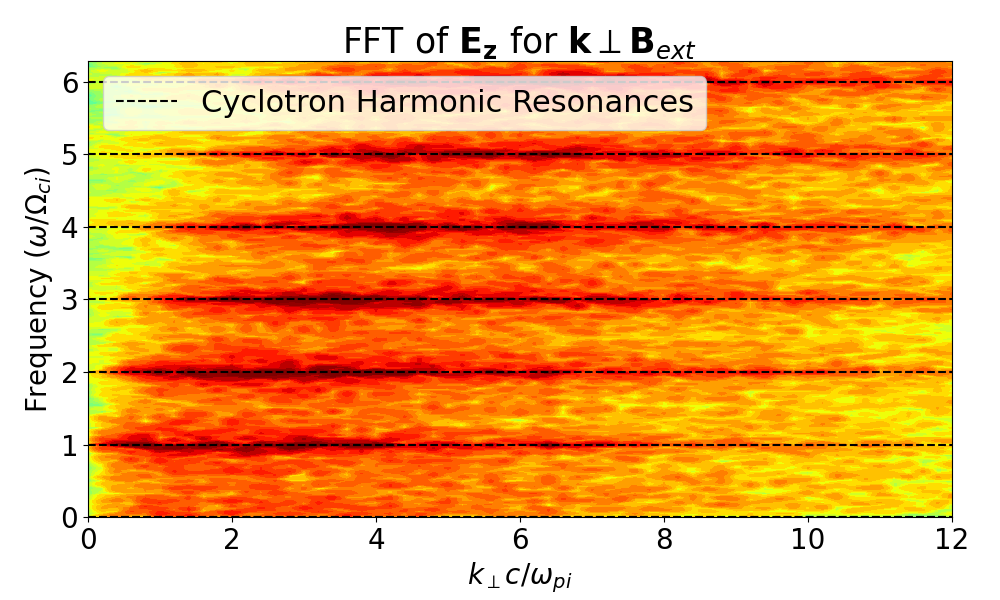}
        \caption{}
        \label{fig:kperp_hot_rkonly_Ez}
    \end{subfigure}
    \caption{Wave spectra of (a) $E_x$, (c) $E_y$ and (e) $E_z(E_\parallel)$ for the `RK-CN-RK' scheme, and of (b) $E_x$, (d) $E_y$ and (f) $E_z(E_\parallel)$ for the fully explicit scheme, for $v_{th,e} = 0.5c$ and $v_{th,i} = 0.25c$, for $\bk \perp \bB_{eq}$. $E_x$ waves are longitudinal and $E_y$ and $E_z(E_\parallel)$ waves are transverse. The dotted lines in (a), (b), (c) and (d) show the Bernstein waves while the dashed lines in (e) and (f) show the ion cyclotron harmonic resonances.}
    \label{fig:kperp_hot}
\end{figure}

We then move on to a hot plasma case with $v_{th,e} = 0.5c$ and $v_{th,i} = 0.25c$. The electric field spectra for this case are shown in the left column of Figure \ref{fig:kperp_hot} for the results obtained using the `RK-CN-RK' time-stepping scheme, and in the right column for the results from the fully-explicit scheme. The results obtained from the `CN-RK-CK' scheme are almost indistinguishable from the `RK-CN-RK' results, and are therefore not shown here. The waves observed in the $E_x$ and $E_y$ spectra are the Bernstein waves described above in Section \ref{sec:perpendicularwaves}. These waves show an excellent agreement with the dotted lines representing the analytical $k-\omega$ curves obtained from equating the first factor of equation \eqref{eq:hot_dispreln_perp} to zero. The $E_z$ spectra clearly show the ion cyclotron harmonic resonances. The theoretical values of the harmonic resonances are marked with dashed lines in Figures \ref{fig:kperp_hot_rkcnrk_Ez} and \ref{fig:kperp_hot_rkonly_Ez}. These harmonic resonances are also observed in the $E_x$ and $E_y$ spectra, but their analytical dashed lines are not shown there for better clarity and to avoid clutter. The numerical time-steps used for each of these schemes are $\Delta t = 0.1$ for `RK-CN-RK', $\Delta t = 0.05$ for `CN-RK-CN', and $\Delta t = 0.01$ for the fully-explicit scheme. The critical time-steps as predicted by the cold plasma stability analysis are $\Delta t \approx 0.85$, $\Delta t \approx 0.42$ and $\Delta t \approx 0.026$ for the `RK-CN-RK', `CN-RK-CN' and fully explicit schemes, respectively. However, for plasmas with significantly high temperatures as considered in this case, the critical time-step is much smaller and difficult to calculate analytically. The time-steps used here were obtained using trial and error. The total computational time is almost the same for the `CN-RK-CN' and `RK-CN-RK' simulations. This is because although the `RK-CN-RK' simulation uses a total number of time-steps that is half of that used by the `CN-RK-CN' simulation, the `RK-CN-RK' scheme has twice the number of LSRK sequences. Besides, although the `CN-RK-CN' scheme uses four times the total number of Crank-Nicolson steps, as compared to the `RK-CN-RK' scheme, the computational cost of these Crank-Nicolson substeps is very small as compared to the RK substeps. The fully-explicit scheme is prohibitively expensive, on account of its much smaller time-step. This results in relative computational savings of $\sim 80\%$, when using either of the IMEX schemes.

\subsection{Damping of Ion Cyclotron Waves}\label{sec:sim_damping}
\begin{figure}
\centering
        \centering
        \includegraphics[width=0.92\linewidth, trim=0pt 0pt 0pt 0pt, clip]{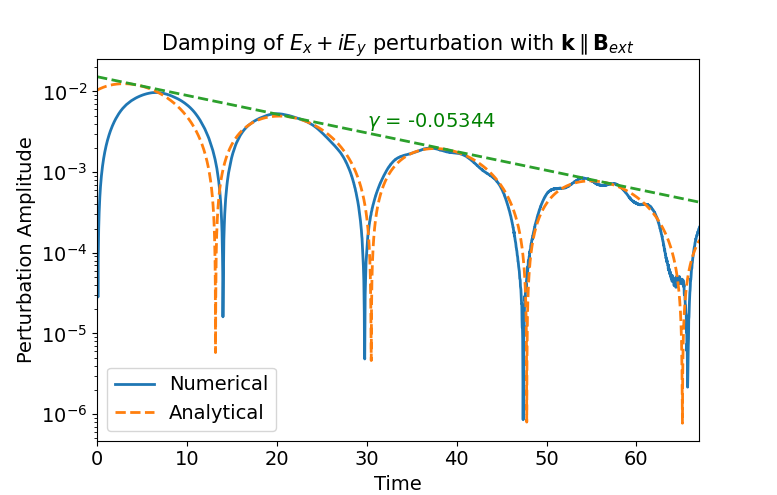}
\caption{Damping of $(E_x + iE_y)$ perturbations for $k_{\parallel} = 1.2$ in a warm plasma with $v_{th,e} = 0.1c$ and $v_{th,i} = 0.05c$. The blue graph shows the result obtained from the numerical simulation, while the dashed orange graph shows the expected analytical result. The dashed green line shows the analytical damping rate.}
\label{fig:Damping}
\end{figure}

The last test we consider is the damping of an ion cyclotron wave in a warm plasma. A single sinusoidal perturbation with a specific wavenumber is initialized in a warm, uniform, equilibrium quasineutral plasma. We consider an ion cyclotron wave with the wavenumber $k_\parallel = 1.2$, traveling in the direction parallel to $\bB_{eq}$. The 256-cell resolution is therefore along this direction. The warm plasma has an electron thermal velocity of $v_{th,e} = 0.1 c$ and ion thermal velocity of $v_{th,i} = 0.05 c$. From the linear perturbation analysis shown in Section \ref{sec:stabilityanalysis}, the following relations between the perpendicular perturbations of the ion current and magnetic field can be derived:
\begin{align}
    \hat{J}_{i,x,k} &= \pm i \hat{J}_{i,y,k}, \label{eq:JixJiyreln} \\
    \hat{B}_{x,k} &= -\frac{\omega}{k_\parallel \Omega_{c,i}} \hat{J}_{i,x,k}, \label{eq:BxJixreln} \\
    \hat{B}_{y,k} &= -\frac{\omega}{k_\parallel \Omega_{c,i}} \hat{J}_{i,y,k}. \label{eq:ByJiyreln}
\end{align}
The `+' sign in equation \eqref{eq:JixJiyreln} corresponds to the ICW branch with the eigenvector $(E_x, -iE_x, 0)$, while the `--' sign corresponds to the CAW-WHW branch with the eigenvector $(E_x, iE_x, 0)$. The dispersion properties of a linear perturbation are obtained by solving the warm plasma dispersion $((\epsilon_{xx}-n^2)(\epsilon_{yy}-n^2)-\epsilon_{xy}\epsilon_{yx}) = 0$ for the waves propagating parallel to the background magnetic field, where $\epsilon_{xx}$, $\epsilon_{yy}$, $\epsilon_{xy}$ and $\epsilon_{yx}$ are given in equations \eqref{eq:epsilonxx_par} and \eqref{eq:epsilonxy_par}. Substituting the wavenumber $k_\parallel = 1.2$, we obtain the solution $\omega \approx 1.657$ corresponding to the CAW-WHW branch, and
$\omega \approx 0.181 - 0.0534i$ corresponding to the ICW branch. Here, the real part is the angular speed and the imaginary part is the damping rate. The high frequency CAW-WHW wave has a zero imaginary component and therefore no damping. Therefore, to observe damping, we initialize the low-frequency ICW wave. The ions are initialized with the velocity perturbation given by:
\begin{gather}\label{eq:vel_perturbation}
  v_x = 0.01 \cos (k_{\parallel} z), \\
  v_y = 0.01 \sin (k_{\parallel} z),
\end{gather}
corresponding to the `+' sign in equation \eqref{eq:JixJiyreln}. Note that the perturbation relations in equations \eqref{eq:JixJiyreln}--\eqref{eq:ByJiyreln} are obtained from a cold plasma stability analysis, and we must therefore use the cold plasma frequency to initialize the magnetic field perturbations. Using the cold plasma ICW branch solution from \eqref{eq:L_ICW}, for the wavenumber $k_\parallel = 1.2$, we obtain $\omega \approx 0.217$. Using this value, the perpendicular magnetic field perturbations are initialized as:
\begin{gather}\label{eq:vel_perturbation}
  B_x = -0.007241 \cos (k_{\parallel} z), \\
  B_y = -0.007241 \sin (k_{\parallel} z),
\end{gather}
according to the relations in equations \eqref{eq:BxJixreln} and \eqref{eq:ByJiyreln}. The resultant damped ion cyclotron wave manifests as sinusoidal perturbations in the electric fields in the perpendicular $x-$ and $y-$ directions. The numerically obtained time-evolution of the $E_x + iE_y$ perturbation amplitude obtained for $\omega(k_\parallel = 1.2)$ is plotted alongside the analytical results in Figure \ref{fig:Damping}. After an initial transient phase, the sinusoidal perturbations match well with the analytical hot plasma ICW frequency and damping rate.

%%%%%%%%%%%%%%%%%%%%%%%%%%%%%%%%%%%%%%%%%%%%%%%%%%%%%%%%%

%%%%%%%%%%%%%%%%%%%%%%%%%%%%%%%%%%%%%%%%%%%%%%%%%%%%%%%%%
\section{Parallelization and Performance}\label{sec:parallelization}
The simulations presented here are performed using the \texttt{GEMPICX} software framework \citep{gempicx}, built upon the \texttt{AMReX} architecture \citep{zhang2021amrex}. \texttt{AMReX} contains functionality for massively parallel, block-structured adaptive mesh refinement (AMR) multiphysics applications, including functionality for particle simulations. Parallelism is achieved primarily through distributed-memory execution using MPI, with additional support for shared-memory threading and accelerator-based backends. The work presented here, however, uses only MPI. \texttt{AMReX} partitions the computational domain into a set of logically rectangular subdomains (grid patches), which are assigned to MPI ranks using a load-balancing strategy aiming to equalize computational work among processes. Each rank is responsible for advancing the solution on its assigned patches, while communication between patches is handled through optimized routines within the framework. Information describing the grid layout is available on all ranks, allowing communication patterns, such as ghost-cell exchanges, to be constructed once and reused efficiently throughout the simulation. Particles are managed using distributed data structures that associate them with the grid patches in which they reside. During time integration, particles may cross patch or process boundaries. When this occurs, they are reassigned to the appropriate destination patch and MPI rank according to their updated positions. This redistribution typically involves communication only between neighboring processes, which helps to keep communication overhead low. Because the method follows a particle-in-cell formulation, frequent coupling between particles and mesh-based fields is required. In practice, this is handled primarily through communication of mesh data. Before interpolating field quantities to particle positions, ghost regions of the mesh are populated using halo exchanges so that each process has access to the necessary field values locally. Following particle-to-mesh deposition, contributions from overlapping regions are combined to ensure consistency and conservation across subdomain boundaries.

 All the large, sparse linear systems in our numerical scheme, i.e. equations \eqref{eq:Eperp_eqn_disc}, \eqref{eq:Epar_eqn_disc}, \eqref{eq:Eperp1_eqn_disc} and \eqref{eq:faraday_disc_E1} are solved in parallel using the \texttt{HYPRE} library \citep{falgout2002, falgout2006design, hypre}. \texttt{AMReX} also provides wrappers to link to the \texttt{HYPRE} library which has functionalities for building and solving these linear systems. In terms of computational cost, particle-related operations dominate the overall runtime, as is generally the case in PIC simulations. These include both particle pushing and the deposition of particle quantities onto the mesh. These particle-based quantities defined on the mesh include charge density, current density, and all the particle-based terms on the right hand side of equation \eqref{eq:Epar_eqn_disc} used to calculate $\arrE_\parallel$. For a simulation with 3000 particles per species per cell, the computational costs of the most dominant computations as percentages of the overall cost are as follows:
\begin{enumerate}
    \item Calculating the right hand side particle contributions in equation \eqref{eq:Epar_eqn_disc} $\approx 41 \%$.
    \item Pushing particles using electromagnetic fields $\approx 26 \%$.
    \item Calculating current density $\approx 18 \%$.
    \item Calculating charge density $\approx 6 \%$.
    \item Redistributing particles $\approx 6 \%$.
    \item Solving linear systems in $\approx 0.4 \%$.
\end{enumerate}
As we reduce the number of particles per cell, the dominance of the particle loop computations reduces. for instance, reducing the particle count to 300 per species per cell, the percentage costs become:
\begin{enumerate}
    \item Calculating the right hand side particle contributions in equation \eqref{eq:Epar_eqn_disc} $\approx 38 \%$.
    \item Pushing particles using electromagnetic fields $\approx 23 \%$.
    \item Calculating current density $\approx 17 \%$.
    \item Calculating charge density $\approx 5.5 \%$.
    \item Redistributing particles $\approx 5 \%$.
    \item Solving linear systems in $\approx 4 \%$.
\end{enumerate}

%%%%%%%%%%%%%%%%%%%%%%%%%%%%%%%%%%%%%%%%%%%%%%%%%%%%%%%%%

%%%%%%%%%%%%%%%%%%%%%%%%%%%%%%%%%%%%%%%%%%%%%%%%%%%%%%%%%
\section{Discussion and conclusions}\label{sec:conclusions}

In this work, we have extended the structure-preserving \texttt{GEMPICX} framework \citep{gempicx} to solve the quasineutral hybrid Vlasov-Maxwell system with drift-kinetic electrons and fully kinetic ions. The resulting numerical method combines the advantages of hybrid modelling with the robustness of structure-preserving discretizations. By deriving the model from a discrete action principle and employing a dual grid mimetic discretization, the resulting scheme inherits key geometric properties of the continuous system, such as the compatibility of Faraday and Amp\`ere equations with the de Rham complex, the solenoidality constraint for the magnetic field, and the quasineutrality constraint on the current density divergence. The formulation avoids explicit time evolution of the discretized electric field. The electric field perpendicular components are instead calculated using the quasineutral Amp\`ere equation, by extracting the $\arrE_\perp-$dependent component of the drift-kinetic electron current. The parallel component $\arrE_\parallel$ is obtained using a curl-curl equation, leading to a well-posed linear system at each time-step. As observed in the dispersion relations and numerical results, this hybrid quasineutral model contains an unphysical, high-frequency wave branch that is numerically stiff, severely restricting the time-step for numerical stability. This restriction was relaxed with the help of implicit-explicit (IMEX) schemes, wherein the stiff term was advanced using an unconditionally stable implicit update, while the rest was advanced explicitly with a much weaker time-step restriction. This approach led to significant computational savings when using the IMEX schemes, as compared to the fully explicit scheme. Our scheme was successfully tested by comparing numerical results with the various waves predicted by dispersion relations, and by simulating the damping of an ion cyclotron wave.

Despite these advancements, our scheme still has several limitations, creating possibilities for improvement in the future. Our model is derived from the gyrokinetic model from \citet{burby2019gauge-free}, by taking the zero Larmor radius limit for the electrons, and also neglecting polarization and magnetization effects. Finite Larmor radius, polarization, and magnetization effects become essential in regimes where electron-scale dynamics become significant. Adding such effects to our model is an interesting avenue for further studies. For the cold plasma case in Section \ref{sec:sim_perpendicularwaves}, our numerical results show observable deviations from the analytical results at high wavenumbers, as seen in Figure \ref{fig:kperp_cold}. These deviations are expected to be mitigated with the use of higher-order Hodge operators that have been found to better capture short-wavelength modes \citep{kormann2024}. Implementing such operators would involve significant changes to the matrices in equations \eqref{eq:Epar_eqn_disc} and \eqref{eq:faraday_disc_E1}, and is left for follow-up efforts. The current work is limited to a Cartesian mesh using slab geometry and periodic boundary conditions. Modelling practical applications involving fusion, space or astrophysical plasmas would provide important benchmarks for our scheme. This would require enhancing the scheme to handle non-periodic boundary conditions \citep{monk2003finite}, and possibly curvilinear meshes \citep{kreeft2011mimetic, MEIERBACHTOL2017796, perse2021}. Additionally, energy-conserving or symplectic time integrators could be explored to further improve long-time stability, compared to standard Runge-Kutta approaches \citep{kraus2017gempic, squire2012geometric, marsden2001discrete}. In PIC methods, statistical noise arising from the finite number of particles per cell introduces numerical errors, which can be particularly problematic when simulating second-order effects such as the ion cyclotron damping process considered in Section \ref{sec:sim_damping}. Control variate ($\delta$f) methods \citep{SONNENDRUCKER2015402, bottino2015monte} offer an effective approach to reducing such noise by evolving only the deviation from a known background distribution, thereby improving accuracy without requiring a prohibitive increase in particle number and computational cost.
%%%%%%%%%%%%%%%%%%%%%%%%%%%%%%%
\section*{Acknowledgments}
Computing resources needed for this work were provided by the Max Planck Computing and Data Facility (MPCDF). The EUROfusion project TSVV-G is acknowledged. This work has been carried out within the framework of the EUROfusion Consortium, funded by the European Union via the Euratom Research and Training Programme (Grant Agreement No 101052200 - EUROfusion). Views and opinions expressed are however those of the author(s) only and do not necessarily reflect those of the European Union or the European Commission. Neither the European Union nor the European Commission can be held responsible for them.

\bibliographystyle{cas-model2-names}
\bibliography{gempic-qn}

\end{document}